\documentclass[electronics,article,accept,pdftex,moreauthors]{Definitions/mdpi} 

\firstpage{1} 
\pubvolume{13}
\issuenum{13}
\articlenumber{2605}
\pubyear{2024}
\copyrightyear{2024}
\makeatletter
\let\c@lofdepth\relax
\let\c@lotdepth\relax
\makeatother
\usepackage{subfigure}
\makeatletter
\renewcommand{\@thesubfigure}{\normalsize(\textbf{\alph{subfigure}})}
\makeatother
\usepackage{soul}
\usepackage{color}
\externaleditor{{Academic Editor:  Fabio Corti}}
\datereceived{30 May 2024} 
\daterevised{23 June 2024} 
\dateaccepted{30 June 2024} 
\datepublished{3 July 2024} 
\hreflink{{https://\linebreak doi.org/10.3390/electronics13132605}} 

\Title{Performance Analysis of Internet of Vehicles Mesh Networks Based on Actual Switch Models} 

\TitleCitation{Performance Analysis of Internet of Vehicles Mesh Networks Based on Actual Switch Models} 

\Author{Jialin Hu 
 $^{1}$, Zhiyuan Ren $^{1,}$*\orcidA{}, Wenchi Cheng $^{1}$, Zhiliang Shuai $^{1}$ and Zhao Li $^{2}$}

\AuthorNames{Jialin Hu, Zhiyuan Ren, Wenchi Cheng, Zhiliang Shuai and Zhao Li}

\AuthorCitation{Hu, J.; Ren, Z.; Cheng, W.; Shuai, Z.; Li, Z.}

\address{%
 $^{1}$ \quad State Key Laboratory of Integrated Services Networks, Xidian University, Xi'an 710071, China; 22011210534@stu.xidian.edu.cn (J.H.); wccheng@xidian.edu.cn (W.C.); 23011210566@stu.xidian.edu.cn (Z.S.)\\
 $^{2}$ \quad Shaanxi Academy of Aerospace Technology Application Company Limited, 
Xi'an 710100, China; zhao.li@lec.ac.cn 
}

\corres{Correspondence: zyren@xidian.edu.cn}

\abstract{\textls[-15]{The rapid growth of the automotive industry has exacerbated the conflict between the complex traffic environment, increasing communication demands, and limited resources. Given the imperative to mitigate traffic and network congestion, analyzing the performance of Internet of Vehicles (IoV) mesh networks is of great practical significance. Most studies focus solely on individual performance metrics and influencing factors, and the adopted simulation tools, such as OPNET, cannot achieve the dynamic link generation of IoV mesh networks. To address these problems, a network performance analysis model based on actual switches is proposed. First, a typical IoV mesh network architecture is constructed and abstracted into a mathematical model that describes how  
		the link and topology changes over time. Then, the task generation model and the task forwarding model based on actual switches are proposed to obtain the real traffic distribution of the network. Finally, a scientific network performance indicator system is constructed. Simulation results demonstrate that, with rising task traffic and decreasing node caching capacity, the packet loss rate increases, and the task arrival rate decreases in the network. The proposed model can effectively evaluate the network performance across various traffic states and provide valuable insights for network construction and enhancement.}}


\vspace{3pt}

\begin{document}




\section{Introduction}

In recent years, with~the development of intelligent devices and wireless communication technology, the~communication distance of devices has gradually increased and has steadily expanded, encompassing diverse geographical terrains and different types of communication services. Numerous communication devices are interconnected, and play an important role in various fields. With~the increase in the global urban population and the rapid development of transportation infrastructures, the~number of vehicles has grown rapidly and become an important means of transport for people to travel. In~this context, recurrent traffic congestion and accidents, traffic inefficiency, and~other problems are becoming increasingly serious. Road safety is now a significant societal issue. Intelligent Traffic System (ITS) establishes a holistic intelligent integrated transportation system by integrating advanced technologies, and~the application of this system is an effective strategy to solve the above problems. Among~them, the~IoV mesh network, as~an important part of ITS, carries substantial responsibility for information exchange and is the key to the operation of the whole system~\cite{I_1}.

In contrast to conventional mobile communication networks, the~IoV mesh network utilizes vehicles with interconnected interface devices as network nodes. According to the different communication objects, the~communication links in vehicular networks include Vehicle-to-Vehicle (V2V) communication and Vehicle-to-Infrastructure (V2I) communication~\cite{I_2}. Among~them, V2V communication facilitates the process of information exchange among vehicles within their communication range. V2I communication realizes the information interaction between vehicles and the Road Side Unit (RSU). V2I communication enables RSUs to offer services like collision warnings, blind spot detection, dynamic route planning, and~other vehicle--road 
 cooperative services. Vehicles 
 in IoV mesh networks have sensors and other devices that can autonomously collect information and transmit it via V2V communication links to enable data interaction between vehicles. Therefore, each node of the vehicle cluster follows the rules of autonomy; cluster members cooperate using information transmission and sharing, showing strong self-organizational capabilities, and can handle tasks collaboratively {\cite{add_2}}. Based on the above analysis, unlike individual vehicles, IoV mesh networks are richer in resources and can handle more complex tasks and provide better services. However, at~the same time, environmental interference, vehicle density and distribution, and~high-speed mobility also pose challenges for scheduling, optimization, and~decision-making of IoV mesh networks~\cite{I_3}. Addressing the bottleneck of collaborative communication modeling in IoV mesh networks is a current research focus in IoV mesh network~construction.

To optimize the problem of IoV mesh network construction, it is firstly necessary to comprehensively analyze the various performances of IoV mesh networks and comprehend how various network parameters affect performance. Among~the factors affecting the performance of IoV mesh networks, the~primary factor is task traffic~\cite{I_4b}. Because~of the high-speed mobility and uneven distribution of vehicles, the~topology in IoV mesh networks rapidly changing, leading to uncertain task forwarding and transmission. With~exceptionally high vehicle density, the~task traffic in the network rises sharply and could potentially result in node cache congestion. Therefore, it is of great practical significance to study the task model and performance metrics system in IoV mesh networks and explore the traffic~distribution.

In addressing vehicular network performance, many researchers have carried out studies on system analysis modeling. For~example, Killat~et~al.~\cite{I_4} assessed performance under the 802.11p standard using four factors: communication distance, node transmission power, packet transmission rate, and~task flow, as~input variables in order to examine the packet reception rate. However, the~stationary nodes in the network are not suitable for the high-speed mobility characteristic of IoV mesh networks. Huang~et~al.~\cite{I_5b} proposed the Real Vehicular Wireless Network model, which allowed for a more realistic capacity analysis in vehicular networks. The~actual geometry of the urban area was represented by a Euclidean planar graph, so as to analyze the interference relations and calculate the asymptotic capacity. Kwon~et~al.~\cite{I_5} represented the network as a geometric structure of straight lines and points for a one-dimensional V2V network with topology changing over time and analyzed network connectivity and capacity using geometric probability. Through mathematical modeling, it was found that the network capacity increases in a specific way, which verifies the validity of the analytical results. Chen~et~al.~\cite{I_6} analyzed the reachable throughput of a vehicular network with finite traffic density under a cooperative communication strategy. By~obtaining a closed-form expression for the reachable throughput, they unveiled the relationship between throughput and key performance-influencing parameters like communication transmission rate and vehicle density. The~above study obtained the expressions for solving the network throughput through mathematical modeling, but did not account for queuing, caching, and~task-forwarding processes at nodes in actual~networks.

Zhang~et~al.~\cite{I_7} built a vehicle movement model to investigate the network coverage problem to ensure stable overall performance. They also compared the performance of network throughput, average network delay, packet loss rate, average routing hops, etc. under different protocols. Saadallah~et~al.~\cite{I_8b} conducted a performance analysis of connectivity and throughput of vehicular networks on motorways, exploring its correlation to vehicle density, number of lanes, etc. The~results, backed up by simulations of realistic vehicular traffic, showed that the V2V + V2I architecture can significantly improve connectivity. Aljabry~et~al.~\cite{I_8} investigated the performance of a common reactive routing protocol where two scenarios are considered. The~first scenario was a comparison between V2V and V2I communication modes. The~second one was made between two maps, Basrah city and the Manhattan grid, in V2V mode. The~implementation of the routing protocol revealed a comparative results analysis using Quality of Service (QoS) parameters, such as network throughput. However, these studies were conducted using prevalent network simulation tools (e.g., OPNET and NS), which cannot accurately address the dynamic link generation problem and portray time-varying features, and~they do not have a suitable methodology to efficiently calculate key performance metrics such as node load rate. Within IoV mesh networks, link construction, disruption, and~recovery among nodes constantly change, and~the performance of individual nodes is also closely related to the network performance. In~addition, based on the actual switches, packet forwarding relies on the port routing forwarding table at the nodes and undergoes a series of complex processes. Therefore, in~order to solve the above problems, this paper proposes a task-forwarding strategy based on the actual switch model, on~the basis of which a network indicator system is constructed to analyze the performance of IoV mesh networks under different vehicle densities, vehicle caching capacities, etc., which lays the foundation for load-balanced routing design, task offloading, and~resource allocation in IoV mesh networks. The~specific research is as follows:\\
\indent (1) We construct 
 a typical network architecture for IoV mesh networks and abstract it into a mathematical model. We create V2V and V2I channel models and vehicle movement models for dynamic link generation and interruption to obtain the network topology over~time.\\
\indent (2) The
 task generation and forwarding model are proposed to obtain the actual traffic distribution of the network, through user traffic modeling and the construction of node port routing tables to establish task scheduling, forwarding, and~transmission strategies based on the actual switch~model.\\
\indent (3) Based
 on the traffic transmission results, a~network performance indicator system is established. We conduct several experimental simulations to assess variations in multiple indicators across diverse vehicle densities and caching capacities, thereby validating the validity and accuracy of the proposed~model.

\section{Related~Works}



Since the pioneering work of Gupta and Kumar in 2000~\cite{I_9}, an~increasing amount of research has focused on the performance of wireless communication networks. As~elements such as the spatial distribution of network nodes, wireless channel fading characteristics, and~other factors influence the interference distribution in wireless networks, current analyses of wireless multi-hop networks primarily concentrate on interference and outage probability, as~well as capacity and throughput assessments. For~example, Vinh~et~al.~\cite{I_10} derived the system outage probability based on a specific fading model for a relayed two-hop collaborative relaying system containing multiple relay nodes for a pair of communication users using amplify-and-forward and decode-and-forward protocols, respectively. Olmedo~et~al.~\cite{I_11b} proposed a wireless TCP protocol improvement considering Negative Acknowledgment (NACK), which allowed for distinguishing between losses due to congestion and losses due to wireless channel problems. This method improved QoS and throughput. Li~et~al.~\cite{I_11} analyzed the outage probability and throughput of a full-duplex orthogonal frequency-division-multiplexing relay network based on the amplify-and-forward protocol and calculated the energy harvesting duration and the signal transmission time, and~the capacity and throughput. The optimal time slot interval between energy harvesting duration and signal transmission time was used as a way to maximize the network~throughput.

However, the~research on wireless communication network performance has generally overlooked the influence of mobility on network layout and node distribution. Consequently, analyzing the performance of IoV mesh networks requires an exploration of the relationship between system parameters and network performance in the context of system characteristics. Rahmani~et~al.~\cite{I_12} conducted a comparison of various network topologies for designing a suitable IP/Ethernet network for audio and video communication in vehicles, considering QoS performance and production costs. Different vehicular networks' capacities were analyzed in the literature~\cite{I_13,I_14,I_15}, with~the capacity determined using the proportionality criterion approach. Distinctions exist among these studies; Nekoui~et~al.~\cite{I_13} assumed all vehicles travel along a single road, while Lu~et~al.~\cite{I_14} assumed that there are multiple roads in the network, but the vehicles can only move around the designated centers. Wang~et~al.~\cite{I_15}  expanded on these scenarios by allowing vehicles to move freely within the network area. By~maximizing the throughput of information transfer between vehicles and RSUs, efficient information transfer path algorithms were derived from this model. Sarvade~et~al.~\cite{I_16} compared the latest IEEE standard 802.11ac with previous MAC protocols based on parameters such as throughput, jitter, and~end-to-end delay using real traffic pattern scenarios. Realistic traffic patterns were generated using SUMO, and~NS-3 evaluated protocol performance. Additionally, Lai~et~al.~\cite{I_17} examined the average packet loss rate's impact under distributed relay selection for multihop broadcasting in vehicular networks, taking into account vehicle mobility, wireless channel conditions, and~media access control. They proposed an analytical model to analyze the average packet loss~rate.

Zhao~et~al.~\cite{I_18} introduced a model for analyzing the QoS and capacity of vehicular networks on one-dimensional motorways and two-dimensional intersection roads. The~validation of the proposed model was conducted through NS-2 simulations and further expanded to include the derivation of additional QoS metrics like packet reception probability, packet acceptance rate, and~broadcast link capacity. Han~et~al.~\cite{I_19b} proposed a distance-weighted back-pressure dynamic routing (DBDR). Simulations in a practical road scenario using NS-2 and VanetMobiSim showed that DBDR outperforms the existing protocols in terms of packet delivery ratio, throughput, and~average packet delay when the network becomes congested. Jiang~et~al.~\cite{I_19b1} constructed a new software-defined networking-based IoV heterogeneous networking measurement framework and proposed a performance measurement and analysis method. The~performance indexes and measurement methods for the loss, throughput, etc. were derived in detail. The~switch selection mechanism and packet sampling process were used to establish optimal measurement points of advantage and quickly obtain the needed measurement information. Wang~et~al.~\cite{I_19} introduced a tree-cut mapping-based average maximum flow solution method (TCMANF) and utilized average network flow (ANF) as a metric to assess the overall network quality of a city. TCMANF quickly determined the maximum flow rate within the inter-node network to obtain the average network flow rate. Zheng~et~al.~\cite{I_20} combined V2V and V2I communication modes to propose a framework for assessing the communication capabilities of vehicular networks operating in mixed traffic scenarios. They introduced a predictive communication strategy to enhance vehicular network capacity in mixed-traffic by pre-caching necessary content in the infrastructure based on predicted vehicle trajectories. Gupta et~al.~{\cite{add_3}} used IEEE 802.11p and IEEE 802.11s for static and moving vehicles working under mesh topology. The~performance evaluation was accomplished by simulation on NS-3. Throughput, packet delivery functions and packet sizes are parameters that have been evaluated. Malnar~et~al. {\cite{add_4}} developed an NS-3-based vehicular network framework to analyze performance metrics such as throughput and packet loss. Several topology-based routing protocols were compared and Expected Transmission Number (ETX) metric was proposed to improve performance. Park~et~al. {\cite{add_5}} designed the Domain-based In-vehicle network Architecture (DIA) and the Zone-based In-vehicle network Architecture (ZIA) for autonomous driving, and~then analyzed and validated the superiority of the ZIA and its suitability as an automotive architecture using the OMNeT++ network simulator.

The above research works investigated the performance of vehicular networks in terms of deriving theoretical expressions and modeling traffic on simulation tools. However, most of them only considered a single metric, and multiple metrics were mostly studied for network capacity and packet loss rate. Due to the high-speed mobility of vehicles in IoV mesh networks and the diversity of user tasks, the~topological connectivity relationships are changing rapidly, and~performance analysis needs to be considered from multiple metrics. In~addition, the~performance of IoV mesh networks is affected not only by vehicle density, which is a factor from the overall perspective of the network, but also by the capacity of the nodes, especially the caching capacity. Furthermore, common simulation tools such as OPNET and NS cannot accurately solve the dynamic generation of links, i.e.,~the time-varying characteristics of links. Tools such as NS-3 do not support metrics other than hop count, and~there is no suitable method to efficiently calculate key performance indicators, such as node load rate {\cite{add_4}}. Therefore, based on the above problems, this paper constructs the task-forwarding model based on actual switches to analyze the network performance in terms of multiple metrics and multiple~factors.

\section{Network Architecture and System~Model}\label{sec3}
\begin{figure}[H]
    \includegraphics[width=10.5cm]{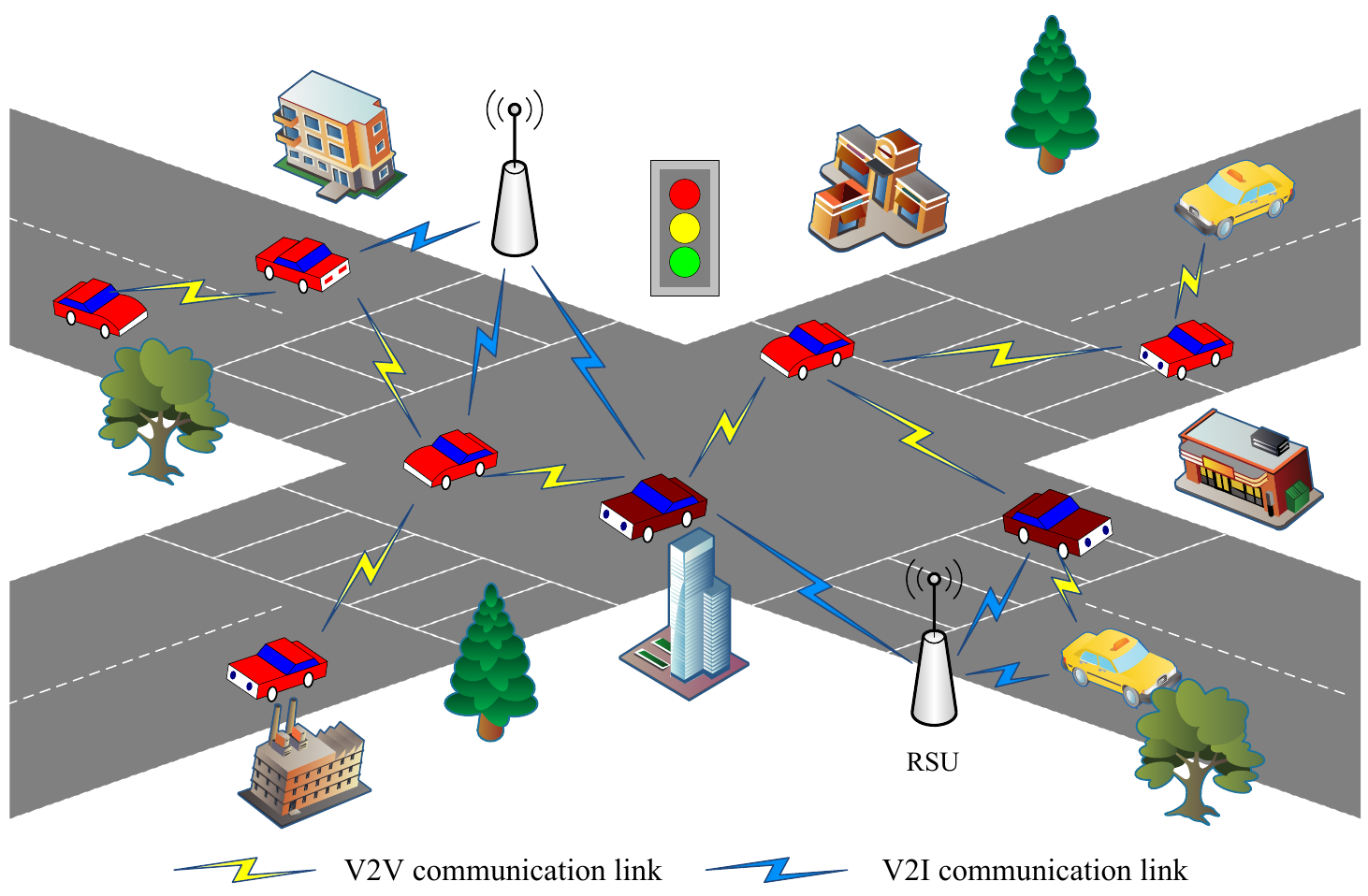}
    \caption{The 
 architecture of IoV mesh~network.}
    \label{fig1}
\end{figure}

The IoV mesh network constitutes a vital component of ITS. Within~this network, each vehicle is equipped with communication and computing devices. Vehicles possess the capability to directly communicate with each other while in motion at high velocities, even in~situations where communication infrastructure such as base stations and wireless access points are unavailable nearby. It is essential to recognize that, within an IoV mesh network, vehicles not only transmit and receive data, but also serve as routers. Consequently, when the sender node is distant from the receiver node, the~sender node transmits data through neighboring nodes in a multi-hop manner. The~IoV mesh network architecture aims to facilitate communication among proximate vehicles, as well as between vehicles and adjacent facilities, which can be divided into two parts: inter-vehicle communication (V2V) and communication between vehicles and infrastructure (V2I). A~typical IoV mesh network architecture is shown in Figure~\ref{fig1}.

\vspace{-0.35cm}

We consider a simplified motion scenario where the newly added vehicles are not involved in the task and, hence, the total number of network nodes remains unchanged for the simulation duration. We set the starting position of the originally existing vehicle that will not leave the simulation area based on its maximum movement speed.

Each vehicle and RSU autonomously generates diverse tasks. For~example, when a vehicle encounters factors such as occlusion or bad weather, and~is unable to make a correct judgment on the current traffic light or the traffic light changes in the coming period, the~RSU can send traffic light information to the neighboring vehicles, and~the vehicle can clarify the traffic light status and forward it to other vehicles after obtaining this information. At~the same time, it can also determine the time for the vehicle to arrive at the intersection according to its own location and the map, so as to realize traffic light speed guidance. Vehicle tasks include augmented reality, real-time video analytics, and~human behavior recognition for drivers, pedestrians, etc.

\textls[-15]{When a vehicle or RSU requires task collection and transmission, it sends requests to neighboring vehicles or RSUs so that different tasks have different destinations. This framework enables vehicles and RSUs to engage in communication, information exchange, collaborative task assignment, and~data processing. Such collaboration allows for real-time environmental sensing and timely adaptation of collaboration strategies across diverse application scenarios. If the network is faulty (e.g., node failure or link outage), there may be an impact on network performance. For~example, when a task has been transmitted to a node, but~that node suddenly fails and is unable to process and forward the task, at~this point, the~packet loss rate of the network will increase, the~task arrival rate will decrease, etc.}

\subsection{Network~Model}

\begin{figure}[H]
    \includegraphics[width=10cm]{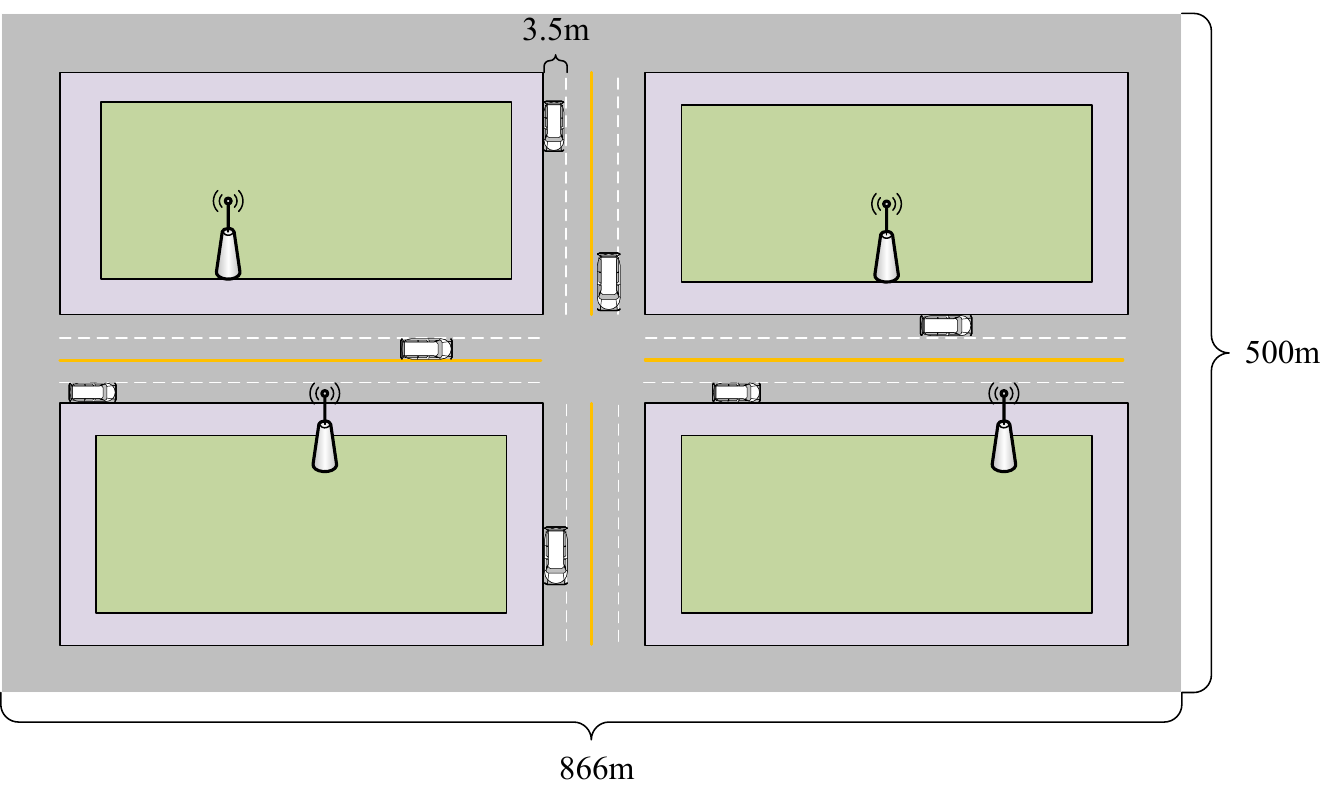}
    \caption{Simulation 
scenario.}
    \label{fig2}
\end{figure}

In order to analyze network performance, this subsection constructs the IoV mesh network dynamic topology. \textls[-15]{Since the topology depends on link construction, the~IoV mesh network channel model is initially established by abstracting the network architecture into a mathematical model. The~scenario involves multiple vehicles and RSUs distributed at an intersection with bi-directional traffic flow. To determine the boundaries of the network model, we refer to the urban case described in 3GPP TR 36.885 {\cite{I_21}}. The~size of the four grids included is 866 m $\times$ 500 m, which has four lanes (two lanes in both forward and reverse directions), and~the lane width is 3.5 m. The~scenario model diagram is shown in Figure~\ref{fig2}.}

\begin{figure}[H]
    \includegraphics[width = 8cm]{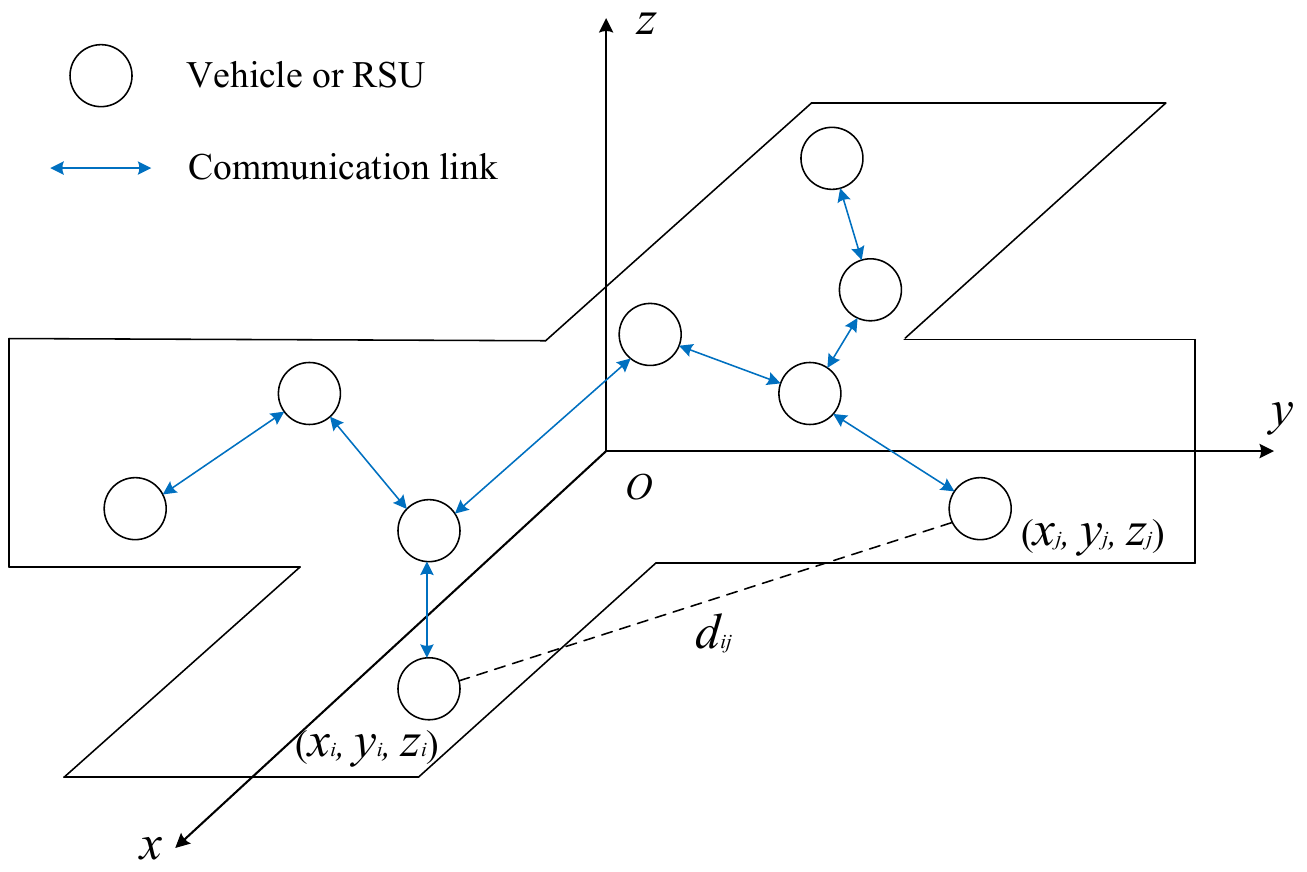}
    \caption{IoV 
 mesh network~model.}
    \label{fig3}
\end{figure}

To facilitate the modeling, a~spatial right-angled coordinate system is established with the center of the road as the coordinate origin, the~\textit{xoy} plane parallel to the ground, and~the z-axis perpendicular to the ground. $U = \{ {u_1},  \cdots ,{u_p}\} $ is defined as the set of vehicles and $S = \{ {s_1},  \cdots ,{s_q}\} $ represents the set of RSUs, where \textit{p} is the total number of vehicles and \textit{q} is the total number of RSUs. The~IoV mesh network model is shown in Figure~\ref{fig3}.

$({x_{i}},{y_{i}},{z_{i}})$ and $({x_{j}},{y_{j}},{z_{j}})$ are the coordinates of any two nodes \textit{i} and \textit{j} in the network, respectively. Therefore, the~Euclidean distance ${d_{ij}}$ between any two nodes at moment \textit{t} can be given by:
\begin{equation}\label{eq:1}
{d_{ij}}(t) = \sqrt {{{({x_i}(t) - {x_j}(t))}^2} + {{({y_i}(t) - {y_j}(t))}^2} + {{({z_i}(t) - {z_j}(t))}^2}}
\end{equation}

According to the destination, the~network has two ways of transmitting data, which are V2I and V2V. When V2V communication is performed, based on existing communication protocols, each vehicle can exchange information with other vehicles located within its maximum access range denoted by \textit{D}. From~this, the~condition that defines the existence of a link between vehicles is shown below, where ${g_{ij}}(t) = 1$ indicates that the link is connected and, vice~versa, interrupted.
\begin{equation}\label{eq:2}
{g_{ij}^{v2v}}(t) = \left\{ {\begin{array}{*{20}{l}}
{1,}&{{d_{ij}^{v2v}}(t) \le {D_j}}\\
{0,}&{otherwise}
\end{array}} \right.
\end{equation}

Therefore, the~link capacity at moment \textit{t} is shown below, where ${B_{ij}}^{v2v}$ and ${h_{ij}}^{v2v}$ are the channel bandwidth and the channel gain between vehicle \textit{i} and vehicle \textit{j}, respectively, and~${{\sigma ^2}}$ denotes the background noise power. Moreover, ${P_{i}}^{v2v}$ denotes the transmit power of vehicle \textit{i}, and~${{A_0}}$ = $-$17.8 dB is a constant parameter.
\begin{equation}\label{eq:3}
R_{ij}^{v2v}(t) = g_{ij}^{v2v}(t) \cdot B_{ij}^{v2v}{\log _2}(1 + \frac{{P_i^{v2v} \cdot h_{ij}^{v2v}}}{{{\sigma ^2} + {A_0}{{(d_{ij}^{v2v}(t))}^{ - 2}}}})
\end{equation}

Similarly, when V2I communication is performed, each RSU can exchange information with vehicles located within its maximum access range. Assuming that the network has been pre-allocated with orthogonal spectrum resources and there is no interference, at~this time, the~link capacity at moment \textit{t} of V2I is shown in Equation~(4). $B_{ij}^{v2I}$ and $h_{ij}^{v2I}$ are the channel bandwidth and the channel gain between vehicle \textit{i} and RSU \textit{j}, respectively \cite{I_22}.
\begin{equation}\label{eq:4}
R_{ij}^{v2I}(t) = g_{ij}^{v2I}(t) \cdot B_{ij}^{v2I}{\log _2}(1 + \frac{{P_i^{v2I} \cdot h_{ij}^{v2I}}}{{{\sigma ^2}}})
\end{equation}

Since this paper focuses on network traffic analysis, assuming all vehicles move in a uniform linear motion for modeling simplicity. A~given time interval \textit{T} is divided into \textit{n} time slots, and~the length of the time slot is $\Delta t = T/n$. The~network topology remains unchanged within each time slot. However, as~vehicles move, connections between vehicles in different time slots change. The~weighted adjacency matrix of the network of the \textit{k} th time slot is defined by Equation~(5), where $\pi _{ij}^k$ represents the link rate. $\pi _{ij}^k = 0$ indicates no connectivity. In~addition, due to changes in the link state, data may be cached on a node during transmission, waiting for link connection, i.e.,~the data caching process.
\begin{equation}\label{eq:5}
{G_k} = \left[ {\begin{array}{*{20}{c}}
0&{\pi _{12}^k}& \cdots &{\pi _{1p}^k}\\
{\pi _{21}^k}&0& \cdots &{\pi _{2p}^k}\\
 \vdots & \vdots & \ddots & \vdots \\
{\pi _{p1}^k}&{\pi _{p2}^k}& \cdots &0
\end{array}} \right]
\end{equation}

Consequently, the~dynamic topology of the IoV mesh network at each moment can be obtained through the above analyses, serving as a foundation for task traffic transmission and performance~analysis.

\subsection{Task Generation~Model}\label{sec3.2}
Since the main service object of the IoV mesh network is vehicle users, and~the network performance will affect the service experience of users, it is significant to analyze quantitatively the quality of user service under different network conditions. For~this reason, this subsection proposes a method for modeling and analyzing traffic with a focus on user experience. First, the~traffic model is constructed, and~the traffic distribution model based on actual switch port forwarding is investigated for analyzing the real-time distribution of data flow within the network, which forms the basis for the subsequent construction of the indicator system and performance~analysis.

\begin{figure}[H]
    \includegraphics[width=11cm]{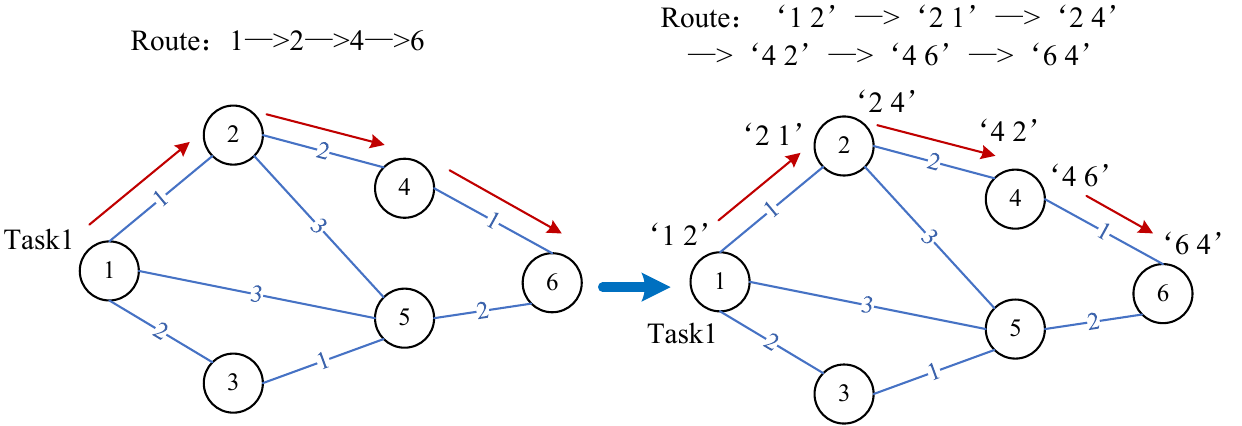}
    \caption{Port 
 route~model.}
    \label{fig4}
\end{figure}

Before modeling the traffic generation of vehicles or RSUs, this subsection constructs the port routing table of each node based on the forwarding strategy of the actual switch, which provides the basis for the generation of task routes. By~traversing the network nodes and utilizing Dijkstra's shortest path algorithm, a~route originating from one node to another network node is created; this route is then converted into a port route to facilitate task forwarding by consulting the node's routing table. This is shown in Figure~\ref{fig4}. In~the topology on the left side of the picture, the~values on the links denote the path distance so that the route for Task 1 can be obtained by the Dijkstra algorithm as 1--->2--->4--->6. 
 In~the topology on the right side of Figure~\ref{fig3}, `1 2' denotes the port utilized by node 1 to communicate with node 2. The~number of ports usually is finite.

\vspace{-0.15cm}

Then, the~traffic generation of vehicles or RSUs is modeled to obtain the task distribution. The~procedure is outlined as follows:
(1) Divide the IoV mesh network into multiple grid spaces and count the number of vehicles and RSUs initiating tasks within each grid.
(2)~Randomly select the destination of each initiating task node.
(3) Construct a traffic model to generate task routes for each initiating task node at each moment, ensuring task QoS values meet the criterion of the average packet loss rate across the network of less than 5${\% }$.
(4) The network exists the high, medium and low priority~tasks.

\subsection{Task Forwarding~Model}\label{sec3.3}
To address the traffic forwarding issue in IoV mesh networks and gain insights into traffic distribution within the network, this subsection proposes a forwarding model based on actual switches to analyze the network traffic in different~states.

\begin{figure}[H]
    \includegraphics[width=8.5cm]{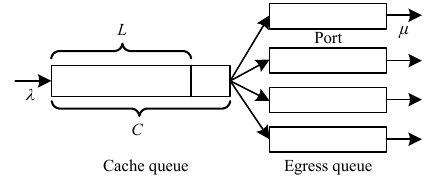}
    \caption{Single-task forwarding~model.}
    \label{fig5}
\end{figure}

Initially, the~analysis is performed for a single service. Within~an IoV mesh network with \textit{p} nodes, the~switch model for forwarding a single task at each vehicle node is shown in Figure~\ref{fig5}. Here, $\lambda $ represents the input traffic of the task at the node, $\mu $ denotes the output traffic (i.e., the~forwarded traffic through the switch), \textit{C} indicates the node's cache capability. The cache refers to the data exchange buffer on the network layer of the node switch, also called the packet buffer size, which is a queuing structure that is used by the switch to coordinate speed-matching problems between different network devices. When a switch does store-and-forward, it presses the packets in the buffer into the egress queue for transmission. \textit{L} denotes the actual cache queue length of the node. Furthermore, the~node's performance incorporates its forwarding capacity, represented by \textit{F}. When the switch forwards, tasks in the cache queue are taken out and pressed into the egress queues of different ports according to rules, and~the egress queue length is the egress link~rate.
\vspace{-3pt}

Based on Section \ref{sec3.2}, the~per-hop port routing of the task can be obtained. Simultaneously, combined with the network topology, we can ascertain the link rate denoted as \textit{R} between the current location node of the task and the subsequent hop node. Consequently, two distinct scenarios emerge for forwarding traffic of a single task:\\
\indent (1) 
 $\lambda  + L \le C$: At this point, the~sum of the actual cache queue length of the task node and the task input traffic is less than or equal to the node's cache capacity. Therefore, the~task has no packet loss on this node, and~the output traffic is constrained by the node's forwarding capacity and the forwarding link rate, i.e.,~$\mu  = \min \{ \lambda  + L,F,R\} $. The~actual cache queue length is further updated as $L' = \lambda  + L - \mu $.\\
\indent (2)  $\lambda  + L > C$: At this point, the~sum of the actual cache queue length of the task node and the task input traffic is greater than the node's cache capacity. So the task will lose some packets at that node, and~the number of packets lost is represented by $Loss = \lambda  + L - C$. In~this case, the~task enters the switch and the packet loss occurs first. Furthermore, the~node's cache is full, and~thus the switch's forwarded traffic is $\mu  = \min \{ C,F,R\} $, and~the actual cache queue length is further updated to be $L' = \lambda  + L - Loss - \mu $.

Once $\mu $ has been obtained, it is necessary to calculate the input traffic at the next hop node. The~input traffic is influenced by the actual link rate denoted as $R'$, as~switches are incapable of sensing interruptions or reductions in link rate (i.e., changes in egress queue lengths) caused by environmental fluctuations. In~conjunction with the changing real network topology, the~following two scenarios exist:\\
\indent (1)  $\mu  \le R'$: At this point, $\mu $ is less than or equal to the number of packets that the link can actually transmit per unit of time, so the number of packets lost on the link is 0, and~the input traffic to the next hop node is $\lambda  = \mu $.\\
\indent (2)  \textls[-15]{$\mu  > R'$: At this point, $\mu $ is greater than the number of packets that the link can actually transmit per unit of time, and~packet loss will occur on the link, with~the number of packets lost being $Loss' = \mu  - R'$. Therefore, the~input traffic to the next hop node is $\lambda  = R' $.}

For the multi-task scenario, when scaling single-task traffic to multi-task traffic, traffic on different ports on the node will be overlapped. An~example of three nodes with two tasks is as follows.

\begin{figure}[H]
    \includegraphics[width=10.5cm]{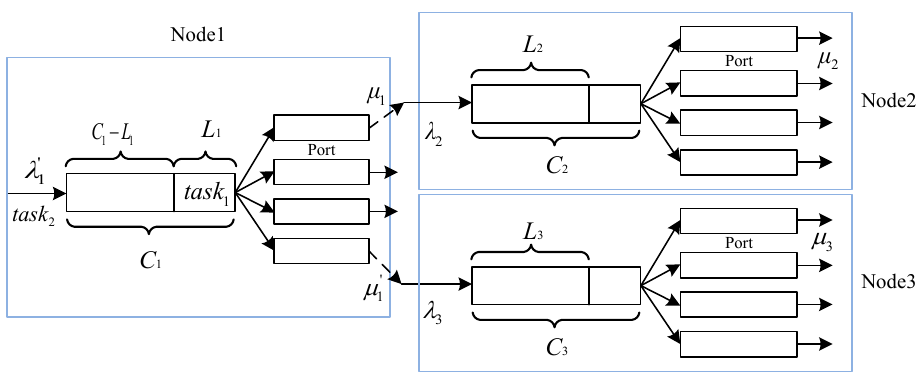}
    \caption{Multi-task forwarding~model.}
    \label{fig6}
\end{figure}

\begin{figure}[H]
    \includegraphics[width=6.5cm]{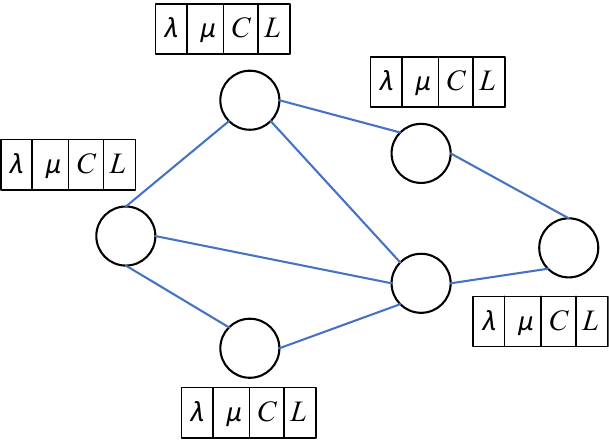}
    \caption{Multi-node 
 multi-task~forwarding.}
    \label{fig7}
\end{figure}

As shown in Figures~\ref{fig6} and \ref{fig7}, at~a specific moment, node 1 has two tasks at the same time, whose F and C are 300 packet/s and 200 packet, respectively. Task 1 has 100 cached packets on the node, and~task 2 is a newly input task with 200 packets. According to the node port routing table and task destinations, it is assumed that the next-hop routes for the two tasks are node 2 and node~3.

At this point, node 1 initiates the sequential arrangement of the two tasks. Tasks are ordered based on a prioritization scheme where high-priority tasks take precedence, followed by medium-priority tasks, then low-priority tasks, while tasks of the same priority are randomized. Then, node 1 assigns forwarding and caching capacities to tasks. \textit{F} and \textit{C} are allocated based on the sorting result; the~more advanced task in the queue is preferred to be allocated with the same number of \textit{F} and \textit{C} as its packets until the allocation is completed. Therefore, the~forwarding and caching sizes of task 1 are 100 and 100, respectively, and~the sizes of task 2 are 200 and 100, respectively. In~cases where multiple tasks are present in the egress link simultaneously, each task is assigned the transmission rate \textit{R} according to the same prioritization principle employed for \textit{F}. Upon~completion of the assignment, the~output flow of each task can be compared and calculated according to the single-task forwarding model, so as to realize the traffic forwarding of multiple~tasks.

 \vspace{-0.15cm}
\unskip

Based on the above analysis, when multiple nodes in the network handle multiple tasks, the~methodology outlined can be employed to compute the input flow, output flow, cache flow, etc., for~each node. This involves directing the traffic from each specific task to its designated forwarding port, allowing for seamless forwarding in conjunction with routing. It ultimately achieves the traffic simulation of the overall network, so as to explore the distribution of the network traffic in different conditions. This approach establishes a groundwork for further investigations into network performance metrics, such as packet loss~rates.

\section{Network Indicator System~Construction}

Due to the intricate physical and logical interconnections across the various levels of the IoV network, consequently, to~effectively analyze the performance of the network, comprehend the task propagation patterns, and~further clarify the relationship between the various levels of the system, it is imperative to establish a comprehensive assessment indication framework. This framework will facilitate the analysis of the IoV network's performance, check the level of congestion in a highly dynamic network, and~enhance the scientific and anticipatory management of IoV~networks.

Based on Section~\ref{sec3}, after~completing the construction of the network traffic transmission model, the~results of the forwarding algorithm can be obtained, and~massive output data can be processed to generate multiple core indicators for the construction of the evaluation system. Given that IoV networks primarily serve users, the~network performance significantly impacts their service experience, which varies according to users' priority. Therefore, from~the task perspective, the~successful task arrival rate, packet loss rate and delay are selected as the core indicators; from the network perspective, when the task is transmitted in the network, the~load rate of the nodes and links are different, and~both of them will indirectly affect user task service experiences. For~example, when the vehicle initiating the service overloads due to failure or excessive link loads between it and the next-hop node, it leads to task accumulation and packet loss. Therefore, using the node load rate, link load rate, and~the actual total task traffic of the network as the core indicators can effectively reflect the blocking situation of the network for performance~evaluation.

In summary, considering that the analysis indicators should be scientific, reasonable and complete, from~the perspective of network structure and performance, this paper adopts five indicators to analyze the network performance. These indices include packet loss rate, task arrival rate, link load rate, node load rate, and~total network~traffic.

\subsection{Packet Loss~Rate}

The packet loss rate is a critical metric for assessing the reliability of network transmission, and serves as an essential indicator of network performance. High packet loss rates indicate network abnormalities. By~tracking the number of packets dropped by each task at every vehicle node, the~packet loss rate of each task can be calculated, leading to the overall packet loss rate for the network. Specifically, assuming that the number of lost packets and the total packets of network tasks are $\{ los{s_1}, \cdots ,los{s_n}\} $, $\{ {\varphi _1}, \cdots ,{\varphi _n}\} $, respectively, where $n$ is the total number of tasks at the current moment, the~overall packet loss rate of the network is:
\begin{equation}
loss\_rate = \frac{{\frac{{los{s_1}}}{{{\varphi _1}}} +  \cdots  + \frac{{los{s_n}}}{{{\varphi _n}}}}}{n}
\end{equation}

\subsection{Task Arrival~Rate}

A series of network issues during data transmission can result in some tasks losing packets or accumulating at nodes, thereby decreasing the success rate of tasks reaching their intended destinations. Therefore, the~successful task arrival rate is a crucial parameter in assessing network performance, as~it reflects the reliability of data transmission. This indicator is calculated as the ratio of packets successfully reaching the destination to the total number of packets sent at the initiation of the task. The~formula is as follows:
\begin{equation}
arrive\_rate = \frac{{\sum {\frac{{{\gamma _i}}}{{{\rho _i}}}} }}{M}
\end{equation}
where ${\gamma _i}$ denotes the total number of packets at which the task $i$ reaches the destination, ${\rho _i}$ denotes the total number of packets at which the task $i$ is initiated, and~$M$ represents the total number of generated tasks. A~lower task arrival rate indicates that most of tasks are congested in the network or packet loss occurs, which will significantly affect the transmission performance of the IoV network and even cause complete network~failure.

\subsection{Node Load~Rate}

The node load rate is defined as the ratio of the total traffic cache and forwarded packets to the node's actual cache capacity. In~scenarios without packet loss at the node, a~lower load rate indicates that tasks are being forwarded to other nodes, thereby increasing node utilization and enhancing overall network performance. However, during~network topology changes or disruptions, the~cache of the node increases, leading to a rise in the node load rate as some tasks may not be promptly forwarded in time. Therefore, this paper introduces the average node load rate to assess the performance of the IoV mesh network.
\begin{equation}
S\_node = \frac{{\sum {\frac{{{\mu _i} + {L_i}}}{{{C_i}}}} }}{N}
\end{equation}

The above equation is the average load rate of the nodes in the network, ${\mu _i} + {L_i}$ denotes the sum of the traffic cache and the number of forwarded packets of the vehicle node $i$, and ${L_i}$ is the cache length at the current moment after updating according to Section~\ref{sec3.3}. ${C_i}$ indicates the actual caching capacity of the vehicle node $i$, and~$N$ represents the total number of nodes in the~network.

\subsection{Link Load~Rate}

During task transmission in a network, traffic flows through the network links. The~ratio of the actual traffic transmitted by a link to the theoretical capacity of the link is defined as the average load rate of the link. This metric provides insight into the utilization of network links; a higher value under no packet loss signifies increased network link utilization. Consequently, this paper introduces the link load rate factor to assess the performance of the IoV mesh network.
\begin{equation}
S\_link = \frac{{\sum {\frac{{{\Psi_{i,j}}}}{{{R_{i,j}}}}} }}{E}
\end{equation}

In the equation above, ${\Psi_{i,j}}$ represents the actual traffic rate of the link from $i$ to $j$, ${R_{i,j}}$ represents the theoretical rate of the link from $i$ to $j$, and~$E$ represents the total number of network links. When the network is disturbed, resulting in packet loss for some tasks, the~load on some links decreases, thereby resulting in performance degradation of the~network.

\subsection{Total Network~Traffic}

The total network task traffic is the aggregate of forwarded task packets and cached task packets within the network. This indicator represents the overall task traffic handled by the IoV mesh network. When the network is congested or disturbed, the~total network task traffic will increase rapidly in a short period of time. Therefore, this paper introduces the total network traffic to evaluate the performance of the IoV mesh network.
\begin{equation}
sumflow = \sum {{\mu _i} + {L_i}}
\end{equation}
where ${\mu _i} + {L_i}$ denotes the sum of traffic caching and forwarding packets of vehicle node $i$. ${L_i}$ in Equation~(10) is the cache length at the current moment after updating as in Equation~(8).

\section{Simulation~Results}
In order to verify the validity of the model proposed in this paper, this subsection performs numerous performance simulations on the IoV mesh network to explore its performance variations under different scenarios. The~simulation platform adopts PyCharm, and~the CPU of the simulation computer is Intel Core i5-1240P. The experiment uses the case shown in Figure~\ref{fig2}. The experiment is set up with four RSUs and 20 vehicles which move uniformly in a linear fashion according to their travel~directions. 

The simulation is divided into 300 time slots, of~which the first 100 time slots are for the upper and lower lanes with the red light on and the left and right vehicles driving through. The~next 50 time slots are for yellow lights for the left and right lanes. The~last 150~time slots have the red light on for the left and right lanes and the up and down vehicles move. Since the vehicle speed is low in the urban scenario, the~effect of the acceleration and deceleration process on the topology change is negligible, and~in order to simplify the system modeling, we simulate the vehicle in uniform linear motion in the moving state. The length of each time slot is 100 ms. Each time slot has a fixed number of vehicles and RSUs to generate tasks. Referring to~\cite{I_21,I_23,I_24,I_25,I_26,I_27,add_6,add_7}, the~rest of the simulation parameters are shown in Table~\ref{tab1}. The maximum packet forwarding rate for vehicle nodes is 10~Gbps {\cite{add_1}}.


\begin{table}[H] 
\caption{System~parameters.\label{tab1}}
\newcolumntype{C}[1]{>{\PreserveBackslash\centering}m{#1}}

\begin{tabularx}{\textwidth}{C{4.5cm}C{1.5cm}C{4cm}C{2cm}}
\toprule
\textbf{Parameter}	& \textbf{Value}	& \textbf{Parameter}	& \textbf{Value}\\
\midrule
V2V link bandwidth & 20 MHz & V2I link bandwidth & 40~MHz \\
Vehicle transmission power & 100 mW & RSU transmission power & 20~dBm \\
Background noise power & $-$100 dBm & Absolute vehicle speed & [20, 40] km/h \\
Maximum communication range of the vehicle & 200 m & RSU coverage range & 500~m \\
Size of vehicle cache & 100 Mb &  Size of RSU cache & 500~Mb \\
\bottomrule
\end{tabularx}
\end{table}
\unskip

\subsection{Network Performance for Different~QoS}\label{sec5.1}

\begin{figure}[H]
\begin{adjustwidth}{-\extralength}{0cm}
  \centering
  \subfigure[~Packet loss rate]
  {\label{fig:subfig1}\includegraphics[width=0.45\linewidth]{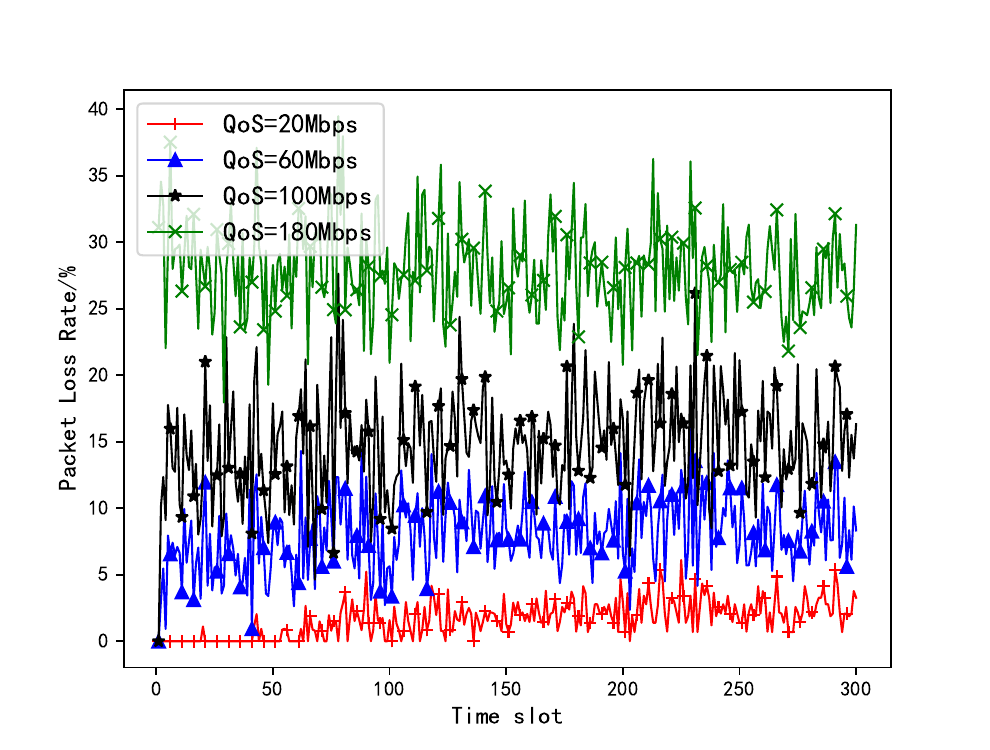}}
  \subfigure[~Task arrival rate]
  {\label{fig:subfig2}\includegraphics[width=0.45\linewidth]{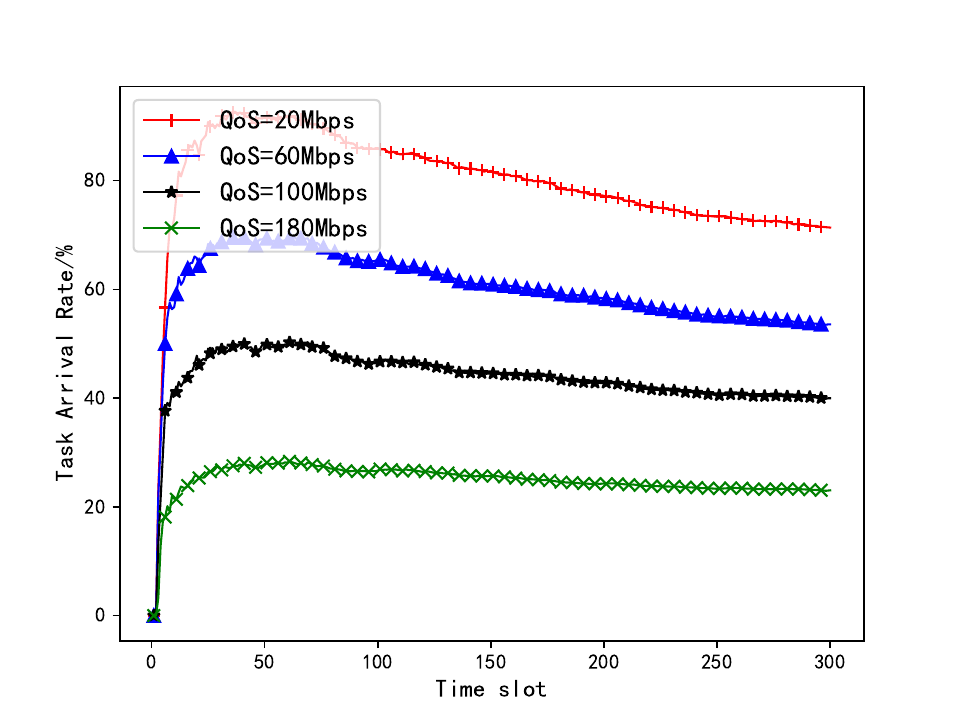}}
\end{adjustwidth}
\caption{Packet loss rate and task arrival rate for different~QoS.}
\label{fig:subfig_1} 
\end{figure}

This subsection focuses on simulating the relationship between task traffic and network performance metrics. The~QoS in the simulation graph represents the traffic of the node initiating the task at each moment. Figure~\ref{fig:subfig_1} explores the fluctuation of network packet loss rate and task arrival rate under different~QoS.

As shown in Figure~\ref{fig:subfig_1}, initially, the~network traffic and task arrival rate exhibit an increasing trend and then change dynamically. This is because under the task QoS and link rate constraints, tasks usually cannot be sent all at once. Consequently, upon~the initiation of a task, network traffic experiences a continuous ascent. Subsequently, task transmission persists over time, with~new tasks entering the network periodically, thereby the traffic distribution on nodes and links is in a dynamic~process.

\vspace{-3pt}

When the QoS is 20 Mbps, the~average packet loss rate is approximately 3${\% }$ and the task arrival rate is around 80${\% }$. As~the QoS increases, the~packet loss rate gradually rises, while the task arrival rate declines. This trend stems from the escalating network ingress traffic resulting in the expansion of the node's cache queue. When the cache queue is full, packet loss occurs. Moreover, the~higher the QoS is, the~easier it is for the cache queue to fill up. In~addition, under~the same QoS setting, the~declining task arrival rate can be attributed to the movement of vehicles causing link disruptions with the change in time, and~most of the packets accumulate at the nodes because they cannot be transmitted according to the original routes when the port routing table is not updated. Simultaneously, the~reduction in task arrival rate is further influenced by continuous task initiation leading to heightened network traffic over~time.

\begin{figure}[H]
\begin{adjustwidth}{-\extralength}{0cm}
  \centering
  \subfigure[~Link load rate]
  {\label{fig:subfig1}\includegraphics[width=0.33\linewidth]{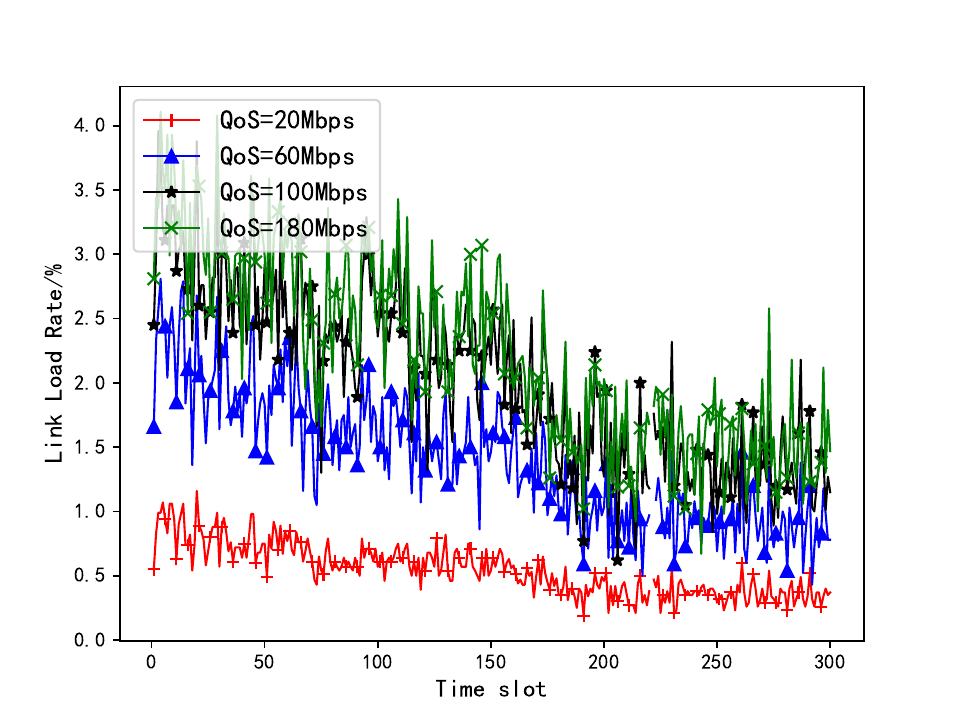}}
  \subfigure[~Total network traffic]
  {\label{fig:subfig2}\includegraphics[width=0.33\linewidth]{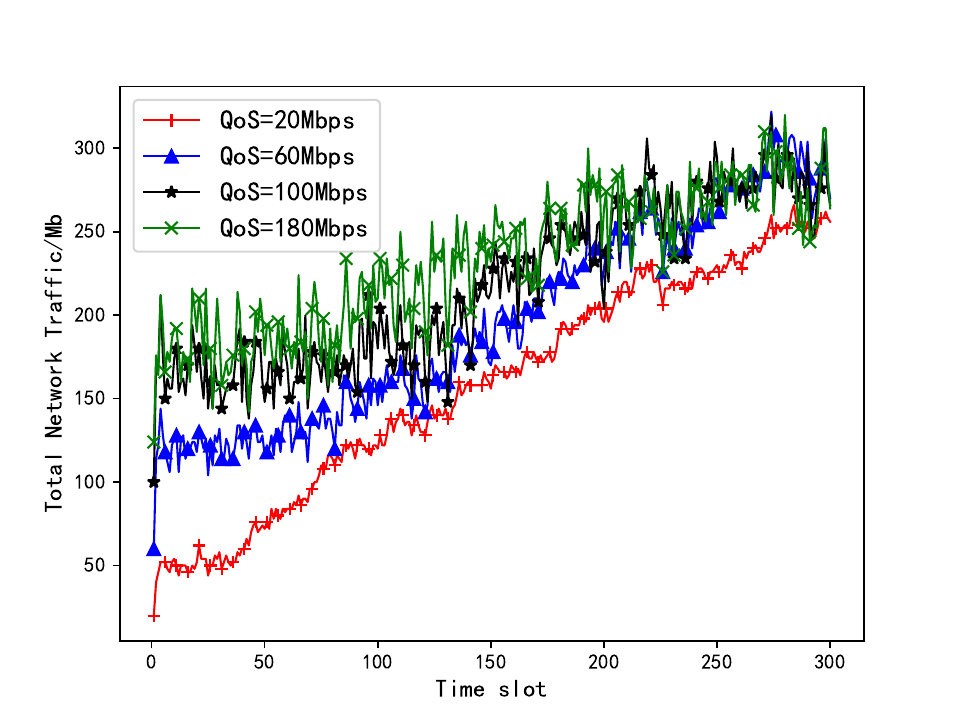}}
  \subfigure[~Node load rate]
  {\label{fig:subfig2}\includegraphics[width=0.33\linewidth]{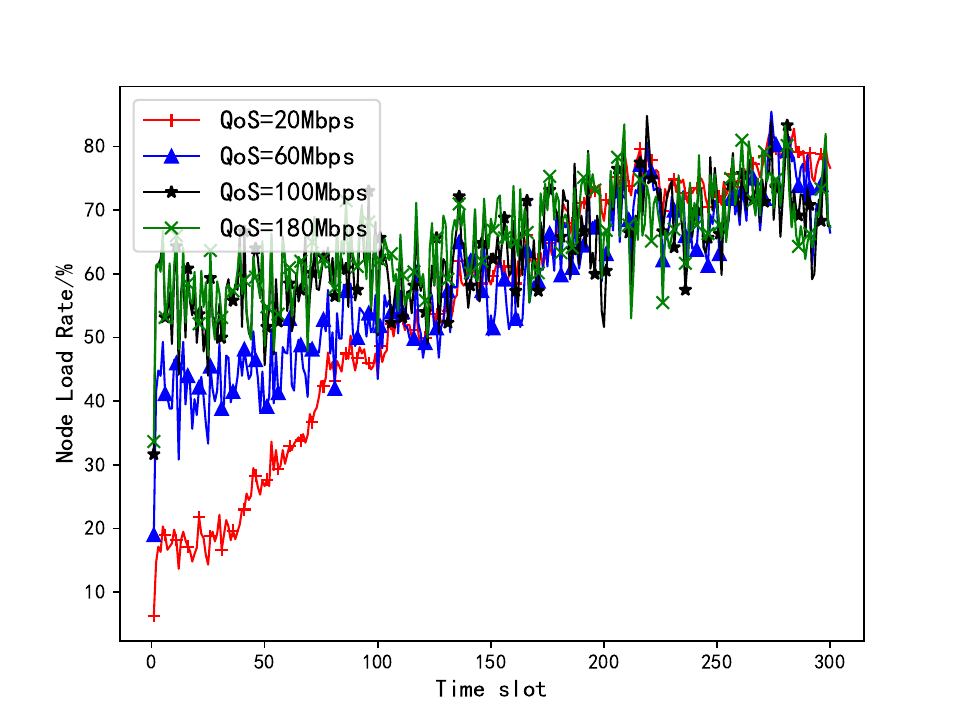}}
\end{adjustwidth}
\caption{Network indicators for different~QoS.}
\label{fig:subfig_2} 
\end{figure}

Figure~\ref{fig:subfig_2} simulates the variations in the other three metrics with QoS. At~lower QoS levels, both the link load ratio and total network traffic exhibit an upward trend with QoS. This is due to the fact that when the QoS is small, the~network loses fewer packets, and~the total traffic increases with the increase in ingress QoS. Simultaneously, the~number of packets forwarded by the nodes increases with the rise in the total number of packets, and~thus the link load rate rises. However, at~higher QoS settings, the~discrepancy between the link load rate and the total network traffic diminishes significantly. This occurrence results from the heightened network packet loss rates in this situation, leading to a small disparity between the number of packets that the nodes can forward after packet loss and the total number of packets carried by the~network.

\vspace{-0.35cm}

\textls[-15]{For the node load rate, the~trend is similar to the other two indicators under higher QoS conditions. The~reason for this is the same as analyzed above when a higher degree of packet loss occurs for both vehicles and RSUs. Conversely, under~lower QoS settings, the~trend in the early stage is similar to the other two indicators. However, in~the later stage, the~cache queue of vehicle nodes tends to reach capacity due to their smaller cache size compared to RSUs. At~this time, a~majority of packet losses in the network occur at vehicle nodes, while the RSUs are minimally affected. Therefore, the~increased total network traffic primarily originates from RSUs, while the increased packet loss from the vehicles results in a decreased overall node load rate in the network at QoS = 60 Mbps compared to QoS = 20 Mbps.}

\subsection{Network Performance for Different Caching~Capacities}\label{sec5.2}

\begin{figure}[H]
\begin{adjustwidth}{-\extralength}{0cm}
  \centering
  \subfigure[~Packet loss rate]
  {\label{fig:subfig1}\includegraphics[width=0.33\linewidth]{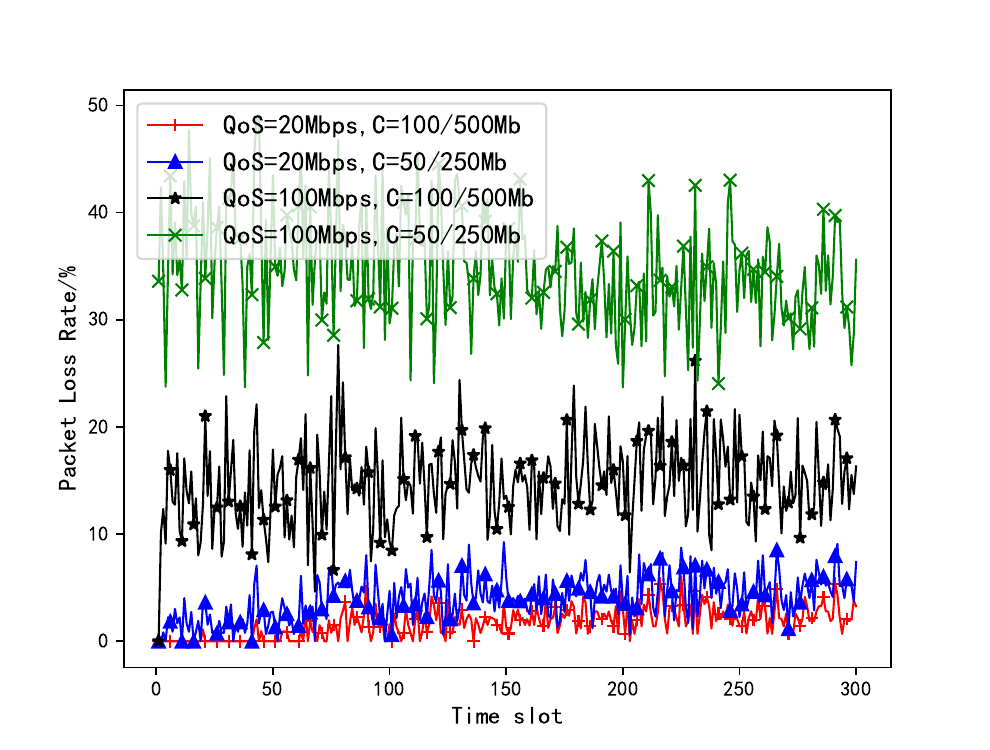}}
  \subfigure[~Task arrival rate]
  {\label{fig:subfig2}\includegraphics[width=0.33\linewidth]{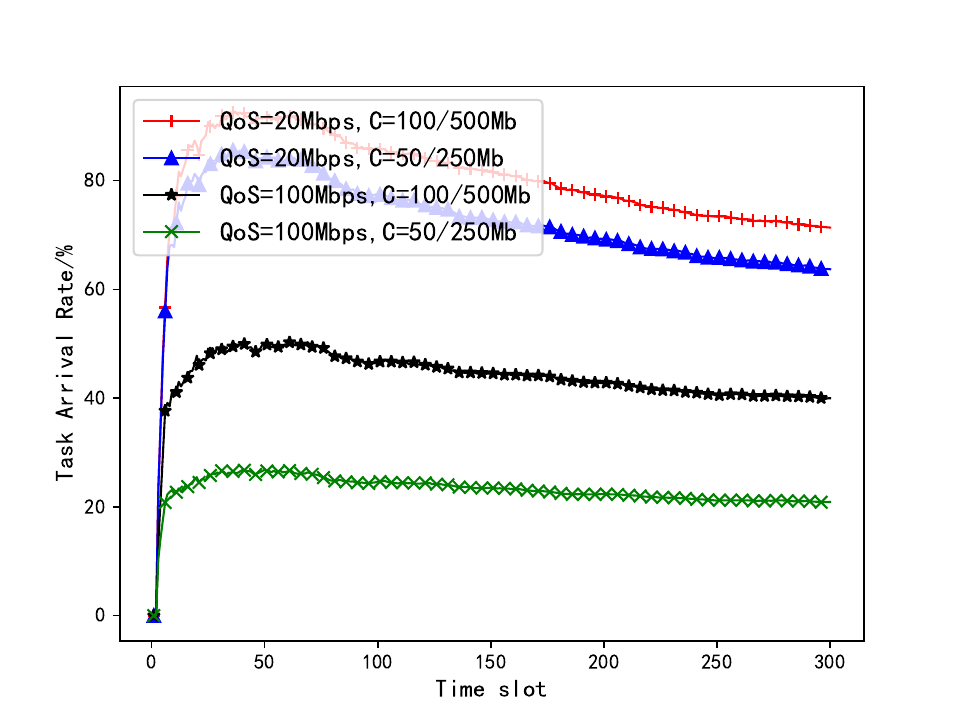}}
  \subfigure[~Link load rate]
  {\label{fig:subfig1}\includegraphics[width=0.33\linewidth]{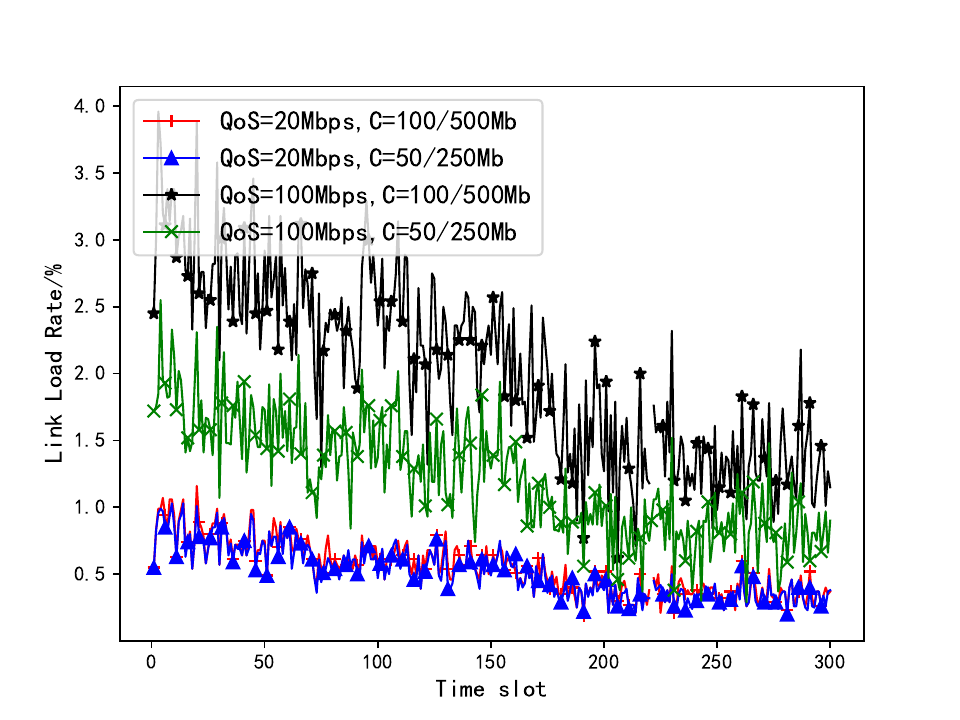}}
  \subfigure[~Total network traffic]
  {\label{fig:subfig2}\includegraphics[width=0.33\linewidth]{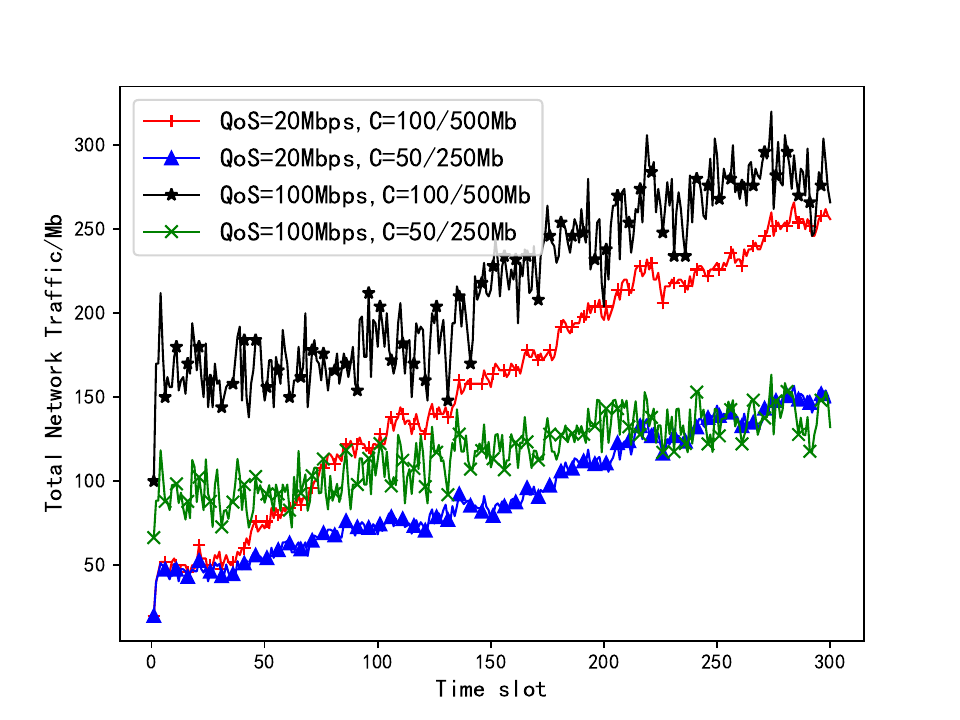}}
\subfigure[~Node load rate]
  {\label{fig:subfig2}\includegraphics[width=0.33\linewidth]{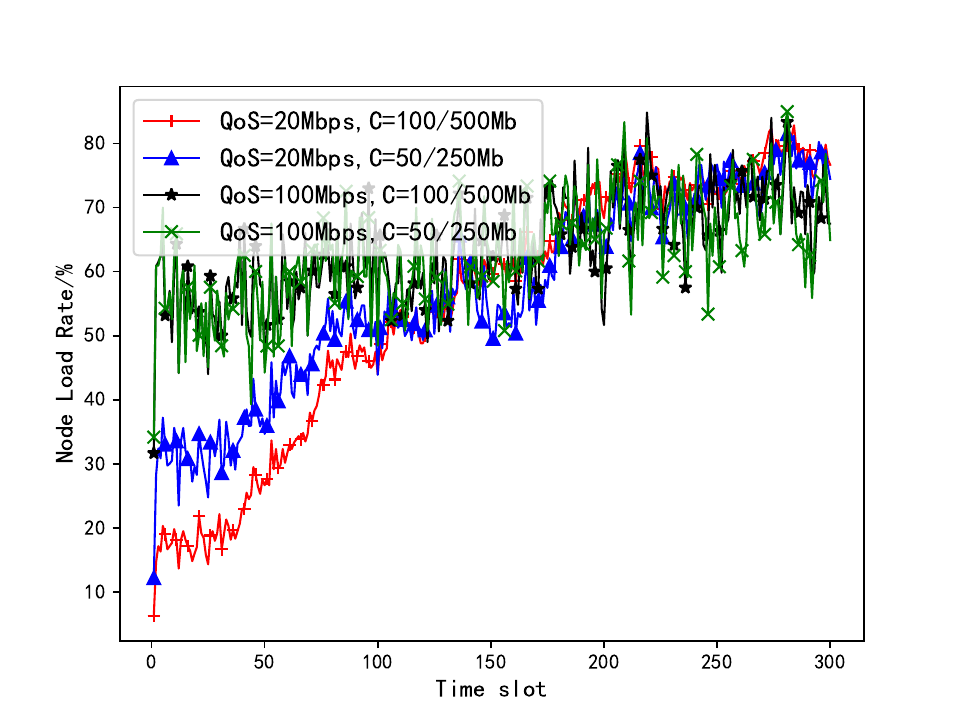}}
\end{adjustwidth}
\caption{Network performance for different caching~capacities.}
\label{fig:subfig_3} 
\end{figure}

From the point of view of the node's performance, Figure~\ref{fig:subfig_3} simulates the variation in network performance with different cache capacities. The~initial vehicle and RSU cache capacities are 100 Mb and 500 Mb, respectively. This subsection reduces both by half and observes the results in conjunction with QoS. From~the perspective of packet loss and task arrival rate, reducing the cache capacity by half has a minor effect on both metrics at low QoS levels, whereas the impact becomes more pronounced at higher QoS values. This is because, in~the case of less network traffic, the~lower cache capacity can carry the task packets efficiently. And,~under high network traffic conditions, the~increase in the total number of packets fills the lower cache queue quickly. Consequently, there is a notable surge in packet loss rate alongside a significant decrease in task arrival~rate.

Regarding the link load rate, reducing the cache capacity by half at the same QoS level reduces this metric. This outcome stems from the decreased storage of packets on the node due to the diminished cache capacity, leading to a corresponding decrease in the number of packets forwarded by the node. Furthermore, the~impact of altering cache capacity on the link load rate is more obvious at QoS = 100 Mbps compared to QoS = 20 Mbps, aligning with the rationale provided for the packet loss rate analysis. Analyzed from the perspective of total network traffic and node load rate, under~equivalent QoS conditions, the~disparity in total network traffic before and after cache capacity adjustment demonstrates a growing trend over time. Moreover, this divergence becomes more conspicuous with higher QoS values. The~reason is that halving the cache capacity leads to a large number of packet losses, the~total number of packets carried by the network decreases, and~the decreasing trend is more obvious with the increase in~traffic.

In addition, the~reason for the trend in node load rate with QoS = 20 Mbps is similar to the analysis in Section~\ref{sec5.1}. When QoS = 100 Mbps, the~total network traffic decreases following a halving of the cache capacity, but~the node load rate remains nearly unchanged at this time. This is because when the cache capacity is larger, although~there are more packets, they are relatively evenly distributed across the nodes. In~contrast, reducing the cache capacity results in vehicle nodes carrying larger packets, so that the overall node load rate of the network does not undergo a significant decrease compared to the previous~one.

\vspace{-0.35cm}


\subsection{Network Performance for Different Vehicle~Densities}

\begin{figure}[H]
\begin{adjustwidth}{-\extralength}{0cm}
  \centering
  \subfigure[~Packet loss rate]
  {\label{fig:subfig1}\includegraphics[width=0.33\linewidth]{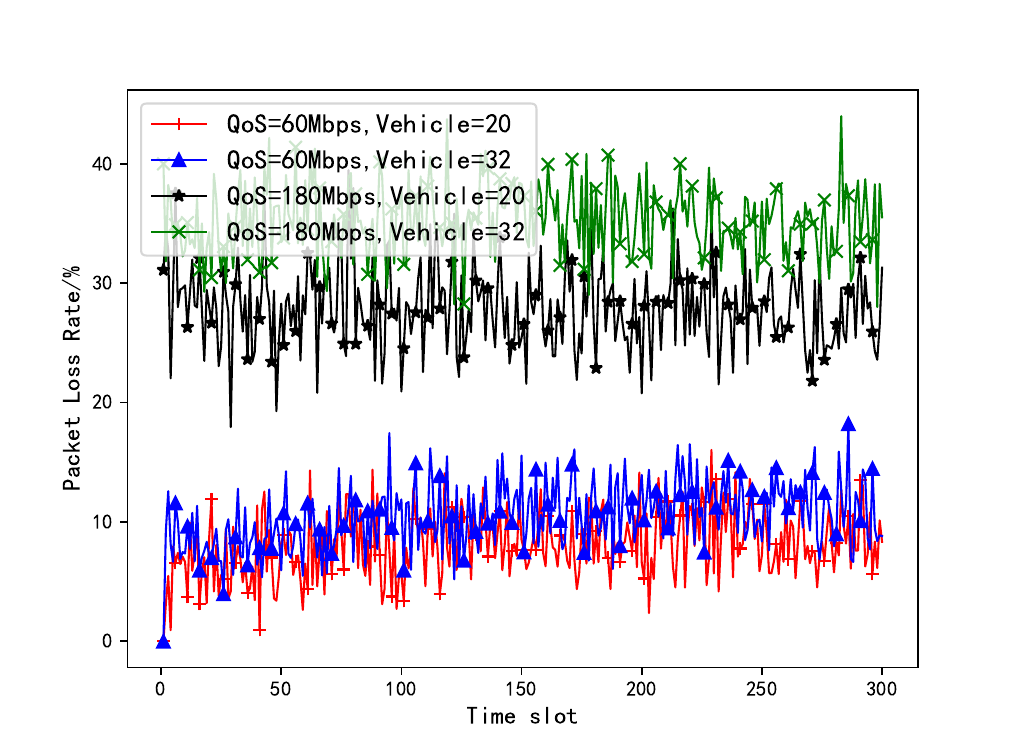}}
  \subfigure[~Task arrival rate]
  {\label{fig:subfig2}\includegraphics[width=0.33\linewidth]{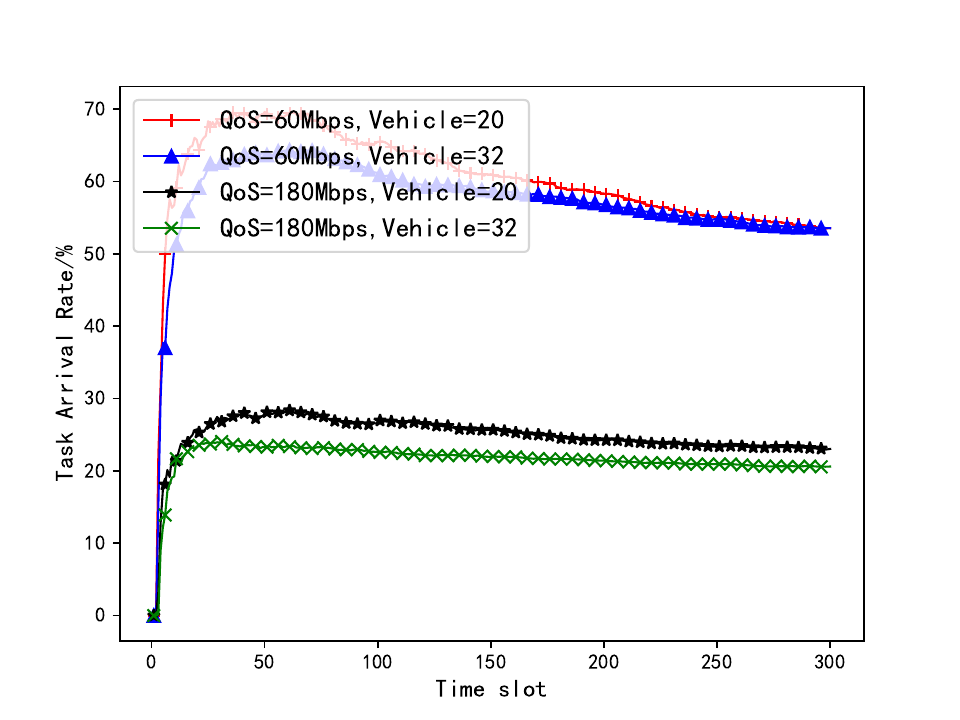}}
  \subfigure[~Link load rate]
  {\label{fig:subfig1}\includegraphics[width=0.33\linewidth]{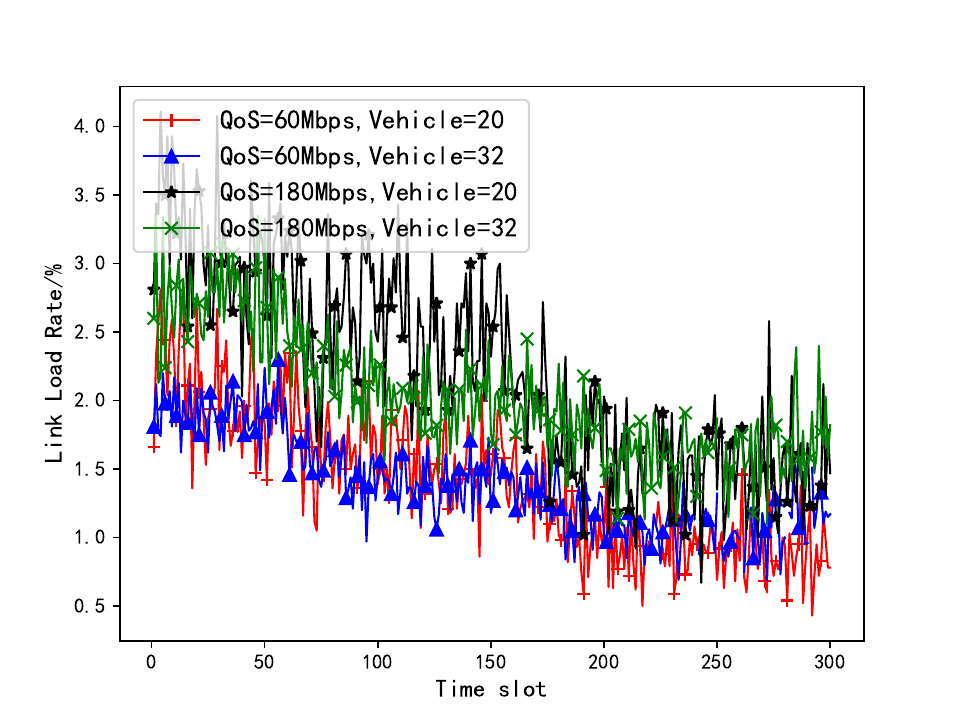}}
  \subfigure[~Total network traffic]
  {\label{fig:subfig2}\includegraphics[width=0.33\linewidth]{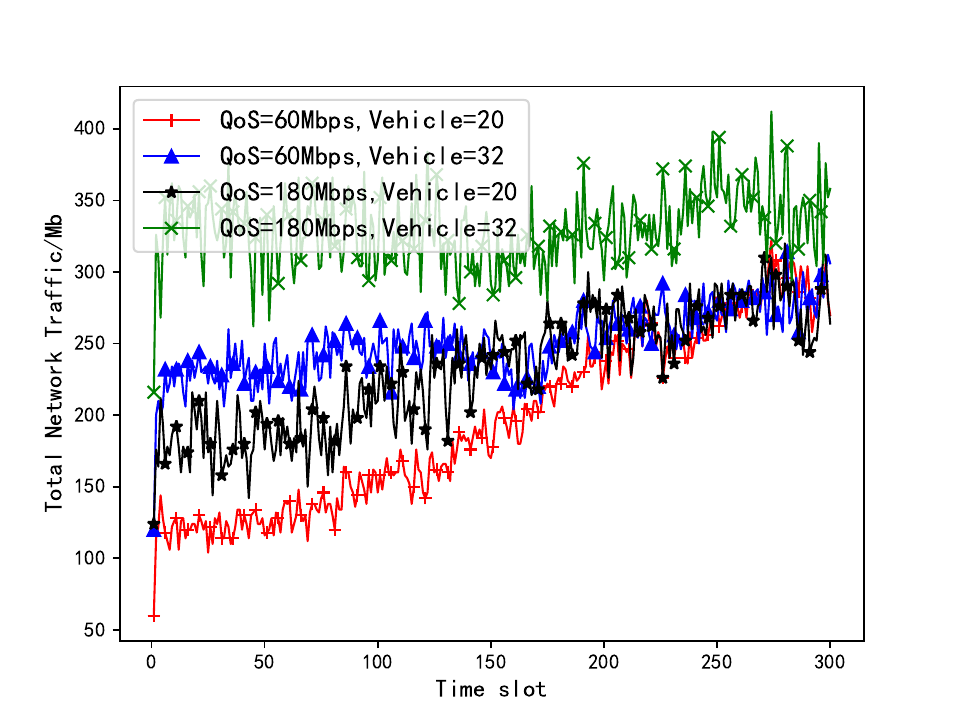}}
\subfigure[~Node load rate]
  {\label{fig:subfig2}\includegraphics[width=0.33\linewidth]{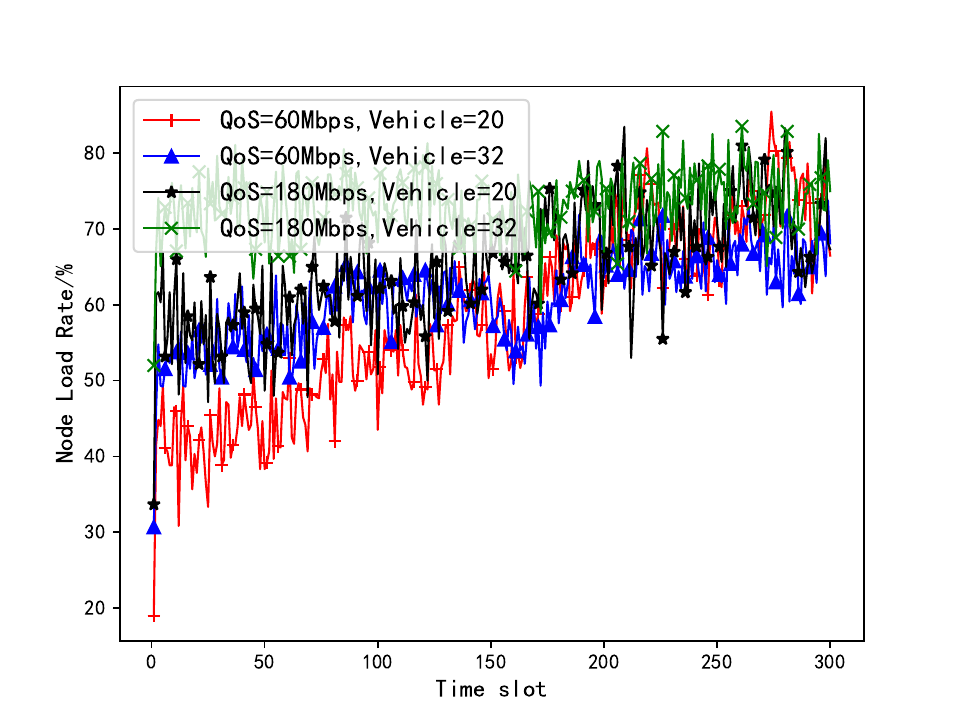}}
\end{adjustwidth}
\caption{Network performance for different vehicle~densities.}
\label{fig:subfig_4} 
\end{figure}

Figure~\ref{fig:subfig_4} simulates the variation in network performance with different numbers of vehicles. As~the quantity and density of vehicles rise, the~number of nodes initiating tasks increases correspondingly, leading to an overall enhancement in the network's total QoS. Consequently, this results in heightened packet loss rates and diminished task arrival rates. However, when QoS = 60 Mbps, the~increase in the number of vehicles has less impact on these two metrics. This phenomenon arises from the network's capacity to efficiently accommodate a greater number of task-initiating nodes under lower ingress QoS levels. In~contrast, at~higher ingress QoS values, the~rapid consumption of limited cache resources by the augmented total packet count hampers network efficiency. As~a consequence, there is a notable surge in packet loss~rates.

Analysis of Figure~\ref{fig:subfig_4} indicates that prior to approximately 150 time slots, networks with a high number of vehicles at the same QoS exhibit a lower link load rate than those with fewer vehicles; however, post 150 time slots, the~pattern reverses. This shift occurs because as vehicle density rises, the~number of links in the network topology also increases correspondingly. According to Equation~(9), this results in a decrease in the link load rate. Conversely, as~time increases, the~impact of an increase in the total number of packets in the network outweighs the effect of an increase in the number of links, leading to a higher volume of packets being forwarded by the nodes and consequently an escalation in the link load rate compared to periods of low vehicle~density.

For the total network traffic metric, the~value of the metric for QoS = 60 Mbps and Vehicle = 32 is higher in the early phase than QoS = 180 Mbps and Vehicle = 20; however, the~values become nearly identical in the later phase. The~reason is that in the early stage of the network, the~network with high vehicle density has enough node caches and links to manage packet storage and forwarding. As~time progresses and the node caches reach capacity, the~disparity diminishes due to packet loss. The~reason for the trend in node load rate is similar to the analysis in Sections~\ref{sec5.1} and \ref{sec5.2}. When QoS = 60Mbps, the~metric displays a decreasing trend in the later stage as the number of vehicles increases. According to Figure~\ref{fig:subfig_4}d and Equation~(8), at~this time, the~total network traffic gap is smaller and the number of nodes is larger, leading to a decrease in the node load~rate.

\vspace{-6pt}

\section{Conclusions}

In this paper, we proposed a network performance analysis model based on actual switches for IoV mesh networks. Initially, a~mathematical model was formulated based on a typical IoV mesh network architecture to obtain the evolving network topology in real-time. Subsequently, by~delving into the ingress and egress traffic forwarding mechanisms of individual nodes and expanding them to a multi-node multi-task environment, we established the task generation and the traffic forwarding model using actual switches. By~integrating the dynamic network topology, we derived the current traffic distribution within the network. This information was then utilized to extract pertinent parameters for constructing the IoV mesh network performance evaluation indicator system, offering a comprehensive and scientific reflection of the network's condition from both the task and network perspective. Simulation results demonstrated that network performance undergoes different but regular changes under diverse ingress QoS levels, caching capacities, and~vehicle densities. The~proposed model provides a rational and effective means of assessing network performance under different traffic~conditions.

In future research, we can study more complex dynamic network scenarios, including networks with diverse node types and multiple channels, based on this network performance analysis model. Additionally, we can also explore problems such as task offloading, resource allocation, and~load-balanced routing design of the network. The~effectiveness and reasonableness of the formulated strategies are assessed by single or multiple~metrics.



\vspace{6pt} 
\clearpage



\authorcontributions{Conceptualization, J.H., Z.R. and W.C.; methodology, J.H., Z.R. and W.C.; software, J.H. and Z.S.; validation, J.H. and Z.S.; formal analysis, J.H.; investigation, J.H.; resources, Z.R.; data curation, J.H. and Z.S.; writing---original draft, J.H.; writing---review and editing, J.H., Z.R., W.C., Z.S. and Z.L.; supervision, Z.R.; project administration, Z.R. and Z.L.; funding acquisition, Z.R. and W.C.; All authors have read and agreed to the published version of the~manuscript.} 

\funding{This research was funded by the Key Research and Development Program of Shaanxi Province of China: 2022ZDLGY05-09.}

\dataavailability{Data are contained within the article.} 




\conflictsofinterest{Author Zhao Li was employed by the company Shaanxi Academy of Aerospace Technology Application Company Limited. The remaining authors declare that the research was conducted in the absence of any commercial or financial relationships that could be construed as a potential conflict of interest. 
} 

\begin{adjustwidth}{-\extralength}{0cm}

\reftitle{References}

\PublishersNote{}
\end{adjustwidth}

\begin{thebibliography}{999}
\bibitem[Author1(year)]{I_1}
Ji, B.; Zhang, X.; Mumtaz, S.; Han, C.; Li, C.; Wen, H.; Wang, D. Survey on the internet of vehicles: Network architectures and applications. {\em IEEE Commun. Stand. Mag.} {\bf 2020}, {\em 4}, 34--41. [\href{http://doi.org/10.1109/MCOMSTD.001.1900053}{CrossRef}]
\bibitem[Author2(year)]{I_2}
Gyawali, S.; Xu, S.; Qian, Y.; Hu, R.Q. Challenges and solutions for cellular based V2X communications. {\em IEEE Commun. Surv. Tutor.} {\bf 2020}, {\em 23}, 222--255. [\href{http://dx.doi.org/10.1109/COMST.2020.3029723}{CrossRef}]

\bibitem[Author8(year)]{add_2}
Ang, L.M.; Seng, K.P.; Ijemaru, G.K.; Zungeru, A.M. Deployment of IoV for smart cities: Applications, architecture, and challenges. {\em IEEE Access} {\bf 2018}, {\em 7}, 6473--6492. [\href{http://dx.doi.org/10.1109/ACCESS.2018.2887076}{CrossRef}]
\bibitem[Author3(year)]{I_3}
Wang, T.H.; Manivasagam, S.; Liang, M.; Yang, B.; Zeng, W.; Urtasun, R. V2vnet: Vehicle-to-vehicle communication for joint perception and prediction. In Proceedings of the European Conference on Computer Vision, Glasgow, UK, 23--28 August 2020; pp. 605--621.

\bibitem[Author3(year)]{I_4b}
Sun, P.; Aljeri, N.; Boukerche, A. Machine learning-based models for real-time traffic flow prediction in vehicular networks. {\em IEEE Netw.} {\bf 2020}, {\em 34}, 178--185. [\href{http://dx.doi.org/10.1109/MNET.011.1900338}{CrossRef}]
\bibitem[Author4(year)]{I_4}
Killat, M.; Hartenstein, H. An empirical model for probability of packet reception in vehicular ad~hoc networks. {\em Eurasip J. Wirel. Commun.} {\bf 2009}, 1--12. [\href{http://dx.doi.org/10.1155/2009/721301}{CrossRef}]

\bibitem[Author4(year)]{I_5b}
Huang, Y.; Chen, M.; Cai, Z.; Guan, X.; Ohtsuki, T.; Zhang, Y. Graph theory based capacity analysis for vehicular ad~hoc networks. In Proceedings of the IEEE Global Communications Conference, San Diego, CA, USA, 6--10 December 2015; pp. 1--5.
\bibitem[Author5(year)]{I_5}
Kwon, S.; Kim, Y.; Shroff, N.B. Analysis of connectivity and capacity in 1-D vehicle-to-vehicle networks. {\em IEEE Trans. Wirel. Commun.} {\bf 2016}, {\em 15}, 8182--8194. [\href{http://dx.doi.org/10.1109/TWC.2016.2613078}{CrossRef}]
\bibitem[Author6(year)]{I_6}
Chen, J.; Mao, G.; Li, C.; Zafar, A.; Zomaya, A.Y. Throughput of infrastructure-based cooperative vehicular networks. {\em IEEE Trans. Intell. Transp.} {\bf 2017}, {\em 18}, 2964--2979. [\href{http://dx.doi.org/10.1109/TITS.2017.2663434}{CrossRef}]
\bibitem[Author7(year)]{I_7}
Zhang, H.; Lu, X. Vehicle communication network in intelligent transportation system based on Internet of Things. {\em Comput. Commun.} {\bf 2020}, {\em 160}, 799--806. [\href{http://dx.doi.org/10.1016/j.comcom.2020.03.041}{CrossRef}]

\bibitem[Author7(year)]{I_8b}
Kassir, S.; Garces, P.C.; de Veciana, G.; Wang, N.; Wang, X.; Palacharla, P. An analytical model and performance evaluation of multihomed multilane VANETs. {\em IEEE/ACM Trans. Netw.} {\bf 2020}, {\em 29}, 346--359. [\href{http://dx.doi.org/10.1109/TNET.2020.3032324}{CrossRef}]
\bibitem[Author8(year)]{I_8}
Aljabry, I.A.; Al-Suhail, G.A. A qos evaluation of aodv topology-based routing protocol in vanets. In Proceedings of the International Conference on Engineering \& MIS, Istanbul, Turkey, 4--6 July 2022; pp. 1--6.

\bibitem[Author8(year)]{I_9}
Gupta, P.; Kumar, P.R. The capacity of wireless networks. {\em IEEE Trans. Inform. Theory} {\bf 2000}, {\em 46}, 388--404. [\href{http://dx.doi.org/10.1109/18.825799}{CrossRef}]

\bibitem[Author8(year)]{I_10}
Vinh, H.D.; Hoang, T.M.; Hiep, P.T. Outage probability of dual-hop cooperative communication networks over the Nakagami-m fading channel with RF energy harvesting. {\em Ann. Telecommun.} {\bf 2021}, {\em 76}, 63--72. [\href{http://dx.doi.org/10.1007/s12243-020-00821-z}{CrossRef}]

\bibitem[Author8(year)]{I_11b}
Olmedo, G.; Lara-Cueva, R.; Martínez, D.; de Almeida, C. Performance analysis of a novel TCP protocol algorithm adapted to wireless networks. {\em Future Internet} {\bf 2020}, {\em 12}, 101. [\href{http://dx.doi.org/10.3390/fi12060101}{CrossRef}]

\bibitem[Author8(year)]{I_11}
Li, J.; Safaei, F. Outage probability and throughput analyses in full-duplex relaying systems with energy transfer. {\em IEEE Access} {\bf 2020}, {\em 8}, 150150--150161. [\href{http://dx.doi.org/10.1109/ACCESS.2020.3016377}{CrossRef}]

\bibitem[Author8(year)]{I_12}
Rahmani, M.; Steffen, R.; Tappayuthpijarn, K.; Steinbach, E.; Giordano, G. Performance analysis of different network topologies for in-vehicle audio and video communication. In Proceedings of the International Telecommunication Networking Workshop on QoS in Multiservice IP Networks, Venezia, Italy, 13--15 February 2008; pp. 179--184.

\bibitem[Author8(year)]{I_13}
Nekoui, M.; Eslami, A.; Pishro-Nik, H. Scaling laws for distance limited communications in vehicular ad~hoc networks. In Proceedings of the IEEE International Conference on Communications, Beijing, China, 19--23 May 2008; pp. 2253--2257.

\bibitem[Author8(year)]{I_14}
Lu, N.; Luan, T.H.; Wang, M.; Shen, X.; Bai, F. Bounds of asymptotic performance limits of social-proximity vehicular networks. {\em IEEE/ACM Trans. Netw.} {\bf 2013}, {\em 22}, 812--825. [\href{http://dx.doi.org/10.1109/TNET.2013.2260558}{CrossRef}]

\bibitem[Author8(year)]{I_15}
Wang, M.; Shan, H.; Luan, T.H.; Lu, N.; Zhang, R.; Shen, X.; Bai, F. Asymptotic throughput capacity analysis of VANETs exploiting mobility diversity. {\em IEEE Trans. Veh. Technol.} {\bf 2014}, {\em 64}, 4187--4202. [\href{http://dx.doi.org/10.1109/TVT.2014.2363791}{CrossRef}]

\bibitem[Author8(year)]{I_16}
Sarvade, V.P.; Kulkarni, S.A. Performance analysis of IEEE 802.11 ac for vehicular networks using realistic traffic scenarios. In Proceedings of the International Conference on Advances in Computing, Communications and Informatics, Udupi, India, 13--16 September 2017; pp. 137--141.

\bibitem[Author8(year)]{I_17}
Lai, W.; Ni, W.; Wang, H.; Liu, R.P. Analysis of average packet loss rate in multi-hop broadcast for VANETs. {\em IEEE Commun. Lett.} {\bf 2017}, {\em 22}, 157--160. [\href{http://dx.doi.org/10.1109/LCOMM.2017.2762686}{CrossRef}]

\bibitem[Author8(year)]{I_18}
Zhao, J.; Wang, Y.; Lu, H.; Li, Z.; Ma, X. Interference-based QoS and capacity analysis of VANETs for safety applications. {\em IEEE Trans. Veh. Technol.} {\bf 2021}, {\em 70}, 2448--2464. [\href{http://dx.doi.org/10.1109/TVT.2021.3059740}{CrossRef}]

\bibitem[Author8(year)]{I_19b}
Han, R.; Guan, Q.; Yu, F.R.; Shi, J.; Ji, F. Congestion and position aware dynamic routing for the internet of vehicles. {\em IEEE Trans. Veh. Technol.} {\bf 2020}, {\em 69}, 16082--16094. [\href{http://dx.doi.org/10.1109/TVT.2020.3041948}{CrossRef}]

\bibitem[Author8(year)]{I_19b1}
Jiang, D.; Wang, Z.; Huo, L.; Xie, S. A performance measurement and analysis method for software-defined networking of IoV. {\em IEEE Trans. Intell. Transp.} {\bf 2020}, {\em 22}, 3707--3719. [\href{http://dx.doi.org/10.1109/TITS.2020.3029076}{CrossRef}]

\bibitem[Author8(year)]{I_19}
Wang, H.; Yang, W.; Wei, W. Efficient algorithms for urban vehicular Ad~Hoc networks quality based on average network flows. {\em Peer-to-Peer Netw. Appl.} {\bf 2024}, {\em 17}, 115--124. [\href{http://dx.doi.org/10.1007/s12083-023-01581-y}{CrossRef}]

\bibitem[Author8(year)]{I_20}
Zheng, Z.; Yue, W.; Li, C.; Duan, P.; Cao, X.; Yue, P.; Wu, J. Capacity of Vehicular Networks in Mixed Traffic with CAVs and Human-Driven Vehicles. {\em IEEE Internet Things} {\bf 2024}, {\em 11}, 17852--17865. [\href{http://dx.doi.org/10.1109/JIOT.2024.3359673}{CrossRef}]

\bibitem[Author8(year)]{add_3}
Gupta, D.; Uppal, A.; Walani, A.; Singh, D.; Saini, A.S. Performance Analysis of Stationary and Moving V2V Communications Using NS3. In \emph{Proceedings of Advances in Smart Communication and Imaging Systems: Select Proceedings of MedCom 2020}; Springer: Singapore, 2021; pp. 475--483.

\bibitem[Author8(year)]{add_4}
Malnar, M.; Jevtić, N. A framework for performance evaluation of VANETs using NS-3 simulator. {\em Promet-Zagreb} {\bf 2020}, {\em 32}, 255--268. [\href{http://dx.doi.org/10.7307/ptt.v32i2.3227}{CrossRef}]

\bibitem[Author8(year)]{add_5}
Park, C.; Park, S. Performance evaluation of zone-based in-vehicle network architecture for autonomous vehicles. {\em Sensors} {\bf 2023}, {\em 23}, 669. [\href{http://dx.doi.org/10.3390/s23020669}{CrossRef}]

\bibitem[Author8(year)]{I_21}
Technical Specification Group Radio Access Network. Study LTE-Based V2X Services, Release 14, Document 3GPP TR 36.885 V14.0.0, 3rd Generation Partnership Project. 2016.  Available online: \url{https://portal.3gpp.org/desktopmodules/Specifications/SpecificationDetails.aspx?specificationId=2934} (accessed on 29 May 2024). 


\bibitem[Author8(year)]{I_22}
Zhang, J.; Guo, H.; Liu, J.; Zhang, Y. Task offloading in vehicular edge computing networks: A load-balancing solution. {\em IEEE Trans. Veh. Technol.} {\bf 2019}, {\em 69}, 2092--2104. [\href{http://dx.doi.org/10.1109/TVT.2019.2959410}{CrossRef}]
%
%
\bibitem[Author8(year)]{I_23}
Al-Hilo, A.; Ebrahimi, D.; Sharafeddine, S.; Assi, C. Vehicle-assisted RSU caching using deep reinforcement learning. {\em IEEE Trans. Emerg. Top. Comput.} {\bf 2021}. [\href{http://dx.doi.org/10.1109/TETC.2021.3068014}{CrossRef}]
%
\bibitem[Author8(year)]{I_24}
Heo, J.; Kang, B.; Yang, J.M.; Paek, J.; Bahk, S. Performance-cost tradeoff of using mobile roadside units for V2X communication. {\em IEEE Trans. Veh. Technol.} {\bf 2019}, {\em 68}, 9049--9059. [\href{http://dx.doi.org/10.1109/TVT.2019.2925849}{CrossRef}]
%
\bibitem[Author8(year)]{I_25}
Zhou, Z.; Liu, P.; Feng, J.; Zhang, Y.; Mumtaz, S.; Rodriguez, J. Computation resource allocation and task assignment optimization in vehicular fog computing: A contract-matching approach. {\em IEEE Trans. Veh. Technol.} {\bf 2019}, {\em 68}, 3113--3125. [\href{http://dx.doi.org/10.1109/TVT.2019.2894851}{CrossRef}]
%
\bibitem[Author8(year)]{I_26}
Ma, B.; Ren, Z.; Cheng, W. Traffic routing-based computation offloading in cybertwin-driven internet of vehicles for v2x applications. {\em IEEE Trans. Veh. Technol.} {\bf 2021}, {\em 71}, 4551--4560. [\href{http://dx.doi.org/10.1109/TVT.2021.3134715}{CrossRef}]

\bibitem[Author8(year)]{I_27}
Hou, X.; Ren, Z.; Wang, J.; Cheng, W.; Ren, Y.; Chen, K.C.; Zhang, H. Reliable computation offloading for edge-computing-enabled software-defined IoV. {\em IEEE Internet Things} {\bf 2020}, {\em 7}, 7097--7111. [\href{http://dx.doi.org/10.1109/JIOT.2020.2982292}{CrossRef}]

\bibitem[Author8(year)]{add_6}
Chang, Q.; Zhang, Z.; Wei, F.; Wang, J.; Pedrycz, W.; Pal, N.R. Adaptive Nonstationary Fuzzy Neural Network. {\em Knowl.-Based Syst.} {\bf 2024}, {\em 288}, 111398. [\href{http://dx.doi.org/10.1016/j.knosys.2024.111398}{CrossRef}]

\bibitem[Author8(year)]{add_7}
Wang, X.; Zhang, K.; Wang, J.; Jin, Y. An enhanced competitive swarm optimizer with strongly convex sparse operator for large-scale multiobjective optimization. {\em IEEE Trans. Evol. Comput.} {\bf 2021}, {\em 26}, 859--871. [\href{http://dx.doi.org/10.1109/TEVC.2021.3111209}{CrossRef}]

\bibitem[Author8(year)]{add_1}
Sun, Q.; Xue, Y.; Li, S.; Zhu, Z. Design and demonstration of high-throughput protocol oblivious packet forwarding to support software-defined vehicular networks. {\em IEEE Access} {\bf 2017}, {\em 5}, 24004--24011. [\href{http://dx.doi.org/10.1109/ACCESS.2017.2767640}{CrossRef}]

%
%
%
%
\end{thebibliography}
\end{document}